# Sources of *E*, *D*, *B*, *H* in electro- and magnetostatics: critical comments

Yu.A. Koksharov

Faculty of Physics, M.V. Lomonosov Moscow State University

Abstract

A critical analysis of teaching approaches to the issue of sources of electrostatic and magnetostatic fields is carried out. Incorrect statements are discussed and their reasons are indicated, in particular, the ambiguity of terminology and going beyond formulas applicability. Unified approach to the study of electromagnetic static field sources based on the Helmholtz theorem is recommended.

# Источники полей *E*, *D*, *B*, *H* в электро- и магнитостатике

Ю. А. Кокшаров

Физический факультет, МГУ имени М.В.Ломоносова

АННОТАЦИЯ

Проведён критический анализ имеющихся в учебной литературе сведений об источниках электростатических и магнитостатических полей. Обсуждаются часто встречающиеся некорректные утверждения и указываются причины, их породившие, в частности, неоднозначность терминологии и выход за рамки применимости используемых формул. Предложено использовать единый подход к изучению источников статических полей $\vec{E}$, $\vec{D}$, $\vec{B}$, $\vec{H}$, основанный на теореме Гельмгольца.

## СОДЕРЖАНИЕ

8. Два вида источников поля $\vec{H}$: токи проводимости и "фиктивные магнитные заряды"

9. Заключительные замечания

Список литературы

**1. Введение**

Найти поле по его источникам - типичная задача электро- и магнитостатики [1]. В некоторых, даже проверенных временем учебниках по электромагнетизму (см, например, [2]-[5]), можно встретить связанные с этой задачей не вполне корректные утверждения, некритичное использование которых может привести к ошибкам. Примеры некоторых таких утверждений:

(У1а) "Единственным источником $\vec{D}$ являются свободные заряды, на которых этот вектор начинается и заканчивается" (см. [2], с.144).

(У1б) "...линии индукции могут начинаться и заканчиваться лишь в тех точках поля, в которых расположены свободные электрические заряды, либо уходить в бесконечность." (см. [3], с.109).

(У2а) "Теорема Гаусса для вектора электрического смещения в диэлектрике имеет такой же вид, как и для напряжённости электрического поля в вакууме. Поэтому все математические соотношения, полученные из неё для вакуума, сохраняют силу и для однородного диэлектрика. Нужно только вектор $\vec{E}$ заменить вектором $\vec{D}$" (см. [4], с.60).

(У2б) "Если однородный диэлектрик заполняет всё пространство, занимаемое полем, то напряжённость $\vec{E}$ поля будет в ε раз меньше напряженности $\vec{E}_0$ поля тех же сторонних зарядов, но в отсутствии диэлектриков" (см. [5], с.78).

(У2в) "...при заданном распределении свободных зарядов потенциал и напряжённость в однородном диэлектрике в ε раз меньше потенциала и напряженности поля в вакууме" (см. [3], с.110).

(У3а) "...при заполнении пространства между проводами однородным магнетиком сила взаимодействия токов возрастает в μ раз" (см. [4], с.254).

(У3б) "одинаковые токи проводимости возбуждают одинаковые напряжённости магнитного поля в вакууме и однородном безграничном магнетике" (см. [2], с.271)

Одна из целей предлагаемых методических заметок − уточнить область применимости процитированных утверждений. Для этого с использованием теоремы Гельмгольца [6]-[8] будут записаны точные общие формулы, связывающие электростатические поля $\vec{E}$, $\vec{D}$ и магнитостатические поля $\vec{B}$, $\vec{H}$ с их источниками. Анализ классификации источников полей (зарядов и токов), используемой в современной учебной и научной литературе, укажет на

"терминологические" причины появления вышеприведённых не вполне точных формулировок. Другая причина возможных недоразумений обусловлена тем, что в преподавании раздела "Электромагнетизм" курса общей физики по историческим причинам сложилась традиция непоследовательного подхода к изучению "вспомогательных" векторов $\vec{D}$ и $\vec{H}$. С одной стороны признаётся их полезность при рассмотрении систем с высокой симметрией, когда очевидно удобство использования теорем Остроградского-Гаусса и Стокса, а также при записи граничных условий. Однако полного представления о векторах $\vec{D}$ и $\vec{H}$, в частности, об их источниках, студенты не получают. Такая ситуация требует исправления.

**2. Историческая справка**

Проблемы, подобные обсуждаемым ниже, не новы. В 30-40-х годах прошлого века в англоязычных физических журналах возникла дискуссия [9]-[12], толчком к которой послужила статья [9], где указывалось на ошибочность утверждения, опирающегося на авторитет Дж.К.Максвелла [13], что напряжённость магнитного поля постоянного магнита в безграничном линейном изотропном магнетике уменьшается (по сравнением с полем этого магнита в вакууме) в μ раз, где μ - магнитная проницаемость магнетика. Дискуссия инициировала создание в 1944 году комиссии "по закону Кулона" (Coulomb's Law Committee), привлекшую ведущих профессоров университетов США. В финальном докладе комиссии [14], [15] содержатся рекомендации по формулировкам и интерпретации фундаментальных законов электромагнетизма для членов американской ассоциации преподавателей физики (American Association of Physics Teachers), анализируются распространённые ошибки в учебниках и способы их избежать.

В докладе комиссии отмечалось, что утверждения в учебниках могут быть неверны ("undoubtedly wrong") по нескольким причинам. Во-первых, авторы учебников могут сами неверно понимать обсуждаемый вопрос. Во-вторых, авторы не очерчивают достаточно чётко область применимости своих утверждений, приписывая частному случаю общий характер. Вторая причина является наиболее распространённой, и авторы доклада советовали, в частности, сделать обязательным правилом сопровождение каждой формулы в учебнике текстом (выделенным курсивом), описывающим область её применимости.

Как показывает наш анализ учебников по электромагнетизму, используемых в современной высшей школе (как российской, так и зарубежной), этот совет по прежнему актуален.

**3. Теорема Гельмгольца и её применение в электростатике и магнитостатике**

Теорема Гельмгольца позволяет записать любое регулярное на бесконечности (спадающее на бесконечности быстрее чем $\sim 1/r$) и дифференцируемое векторное поле $\vec{K}$ в виде интегралов по всему пространству от его дивергенции и ротора [7].

$$\vec{K}(x,y,z) = \frac{1}{4\pi}\{\nabla \times \int_\infty \frac{\nabla \times \vec{K}}{r} dV' - \nabla \int_\infty \frac{(\nabla' \vec{K})}{r} dV'\} \qquad (1)$$

где $r$ - расстояние от элемента интегрирования d$V'$ с координатами ($x'$, $y'$, $z'$)' до точки с координатами ($x$, $y$, $z$), в которой вычисляется поле $\vec{K}$. Оператор $\nabla$ применяется к координатам ($x$, $y$, $z$), оператор $\nabla'$ к координатам ($x'$, $y'$, $z'$)'.

Эта теорема может быть применена к электрическим и магнитным векторам $\vec{E}$, $\vec{D}$, $\vec{B}$, $\vec{H}$, если заряды и токи находятся в ограниченной области пространства.

Формула (1) демонстрирует, в частности, важность уравнений Максвелла, в левой части которых мы видим дивергенцию или ротор, а правая часть содержит плотность зарядов или токов, или равна нулю:

div($\vec{E}$)=$\rho/\varepsilon_0$ \qquad (2)

rot($\vec{E}$)=0 \qquad (3)

div($\vec{B}$)=0 \qquad (4)

rot($\vec{B}$)=$\mu_0 \vec{j}$ \qquad (5)

где $\rho$ - объёмная плотность полного электрического заряда, $\vec{j}$ - плотность полного тока. Отметим, что $\frac{\partial \vec{B}}{\partial t} = 0$ и $\frac{\partial \vec{E}}{\partial t} = 0$ в рассматриваемом статическом случае.

Для векторов $\vec{D}$, $\vec{P}$ и $\vec{H}$, $\vec{M}$ также можно записать формулы для их дивергенций и роторов, которые в статическом случае имеют вид:

div($\vec{D}$)=$\rho_f$ \qquad (6)

div($\vec{P}$)=$-\rho_b$ \qquad (7)

rot($\vec{D}$)=$-$rot($\vec{P}$) \qquad (8)

rot($\vec{H}$)=$\vec{j}_c$ \qquad (9)

rot($\vec{M}$)=$\vec{j}_M$ \qquad (10)

div($\vec{H}$)=$-$div($\vec{M}$) \qquad (11)

где $\vec{P}$ – вектор поляризации диэлектрика, $\vec{M}$ – вектор намагниченности магнетика, $\rho_f$ и $\rho_b$ – объёмная плотность свободного и связанного электрического заряда, соответственно; $\vec{j}_c$ и $\vec{j}_M$ – плотность тока проводимости и тока намагничивания, соответственно.

С учетом (2) и (3) формула (1) для вектора $\vec{E}$ примет вид:

$$\vec{E}(x,y,z) = -\frac{1}{4\pi\varepsilon_0}\nabla\int_\infty \frac{\rho(x',y'z')}{r}dV' \qquad (12)$$

где ρ(x',y',z') - объёмная плотность полного электрического заряда,

$$r = \sqrt{(x-x')^2 + (y-y')^2 + (z-z')^2}$$

- модуль вектора $\vec{r}$, направленного от точки (x', y', z)' к точке (x, y, z).

Если использовать векторное тождество

$$\nabla(\frac{1}{r}) = -\frac{\vec{r}}{r^3} \qquad (13)$$

то меняя местами дифференцирование и интегрирование в формуле (12), можно преобразовать её к известному выражению для напряжённости электростатического поля, обычно выводимому из закона Кулона и принципа суперпозиции:

$$\vec{E}(x,y,z) = \frac{1}{4\pi\varepsilon_0}\int_\infty \frac{\rho(x',y'z')\vec{r}}{r^3}dV' \qquad (14)$$

Интегрирование по всему пространству в (14) можно заменить интегрированием по ограниченному объёму Ω, в котором расположены заряды, и по поверхности Σ, ограничивающий этот объём.

$$\vec{E}(x,y,z) = \frac{1}{4\pi\varepsilon_0}\{\int_\Omega \frac{\rho(x',y'z')\vec{r}}{r^3}dV' + \int_\Sigma \frac{\sigma(x',y'z')\vec{r}}{r^3}dS'\} \qquad (15)$$

где σ(x',y',z') - поверхностная плотность полного электрического заряда.

Согласно (6)-(8) формула (1) для вектора $\vec{D}$ должна содержать два слагаемых:

$$\vec{D}(x,y,z) = \frac{1}{4\pi}\int_\infty \frac{\rho_f(x',y',z')\vec{r}}{r^3}dV' + \frac{1}{4\pi}\nabla\times\int_\infty \frac{\vec{j}_P(x',y',z')}{r}dV' \qquad (16)$$

где введено обозначение

$$\vec{j}_P = \text{rot}(\vec{P}) \qquad (17)$$

для "фиктивного поляризационного тока" (ФПТ) $\vec{j}_P$. Это вспомогательное (не описывающее реальный электрический ток) понятие позволяет [8] использовать в задачах электростатики математическую аналогию с формулами магнитостатики, содержащими реальный ток намагничивания $\vec{j}_M$ (10).

Второе слагаемое в (16) можно преобразовать, используя формулу для ротора произведения скалярного и векторного полей:

$$\nabla\times\frac{\vec{j}_P(x',y',z')}{r} = \frac{\nabla\times\vec{j}_P(x',y',z')}{r} + \frac{\vec{j}_P(x',y',z')\times\vec{r}}{r^3} = \frac{\vec{j}_P(x',y',z')\times\vec{r}}{r^3} \qquad (18)$$

В результате (16) принимает вид:

$$\vec{D}(x,y,z) = \frac{1}{4\pi} \int_\infty \frac{\rho_f(x',y',z')\vec{r}}{r^3} dV' + \frac{1}{4\pi} \int_\infty \frac{\vec{j}_P(x',y',z') \times \vec{r}}{r^3} dV' \qquad (19)$$

Переходя к интегрированию по конечной области пространства $\Omega$ и её поверхности $\Sigma$, получаем:

$$\vec{D}(x,y,z) = \frac{1}{4\pi} \{ \int_\Omega \frac{\rho_f(x',y',z')\vec{r}}{r^3} dV' + \int_\Sigma \frac{\sigma_f(x',y',z')\vec{r}}{r^3} dS' + \int_\Omega \frac{\vec{j}_P(x',y',z') \times \vec{r}}{r^3} dV' + \int_\Sigma \frac{\vec{i}_P(x',y',z') \times \vec{r}}{r^3} dS' \} \quad (20)$$

где $\vec{i}_P = \vec{n} \times \Delta\vec{P}_\tau$ \qquad (21)

– линейная плотность поверхностного ФПТ, $\vec{n}$ – единичный вектор внешней нормали к $\Sigma$, $\Delta\vec{P}_\tau$ – изменение тангенциальной компоненты вектора поляризации при переходе из $\Omega$ наружу через поверхность $\Sigma$, $\sigma_f$ – поверхностная плотность свободного электрического заряда.

Для магнитных векторов $\vec{B}$ и $\vec{H}$, проводя аналогичные преобразования уравнения (1) с учётом (4), (5), (9)-(11), можно получить формулы ([8]):

$$\vec{B}(x,y,z) = \frac{\mu_0}{4\pi} \int_\infty \frac{\vec{j}(x',y',z') \times \vec{r}}{r^3} dV' \qquad (22)$$

$$\vec{H}(x,y,z) = \frac{1}{4\pi} \int_\infty \frac{\rho_M(x',y',z')\vec{r}}{r^3} dV' + \frac{1}{4\pi} \int_\infty \frac{\vec{j}_c(x',y',z') \times \vec{r}}{r^3} dV' \qquad (23)$$

где $\vec{j}(x',y',z') = \vec{j}_M(x',y',z') + \vec{j}_c(x',y',z')$ - плотность полного тока,

равная сумме плотности токов намагничивания $\vec{j}_M(x',y',z')$ и проводимости $\vec{j}_c(x',y',z')$,

$$\rho_M(x',y',z') = -\mathrm{div}(\vec{M}) \qquad (24)$$

– объёмная плотность "фиктивного магнитного заряда" (ФМЗ), вводимого, как и ФПТ $\vec{j}_P$ с целью использования математической аналогии между уравнениями электро- и магнитостатики.

Так же как в электростатическом случае, в (22) и (23) можно перейти к интегрированию по конечному объёму $\Omega$ и ограничивающей его поверхности $\Sigma$:

$$\vec{B}(x,y,z) = \frac{\mu_0}{4\pi} \{ \int_\Omega \frac{\vec{j}(x',y',z') \times \vec{r}}{r^3} dV' + \int_\Sigma \frac{\vec{i}(x',y',z') \times \vec{r}}{r^3} dS' \} \qquad (25)$$

$$\vec{H}(x,y,z) = \frac{1}{4\pi} \{ \int_\Omega \frac{\vec{j}_c(x',y',z') \times \vec{r}}{r^3} dV' + \int_\Sigma \frac{\vec{i}_c(x',y',z') \times \vec{r}}{r^3} dS' + \int_\Omega \frac{\rho_M(x',y',z')\vec{r}}{r^3} dV' + \int_\Sigma \frac{\sigma_M(x',y',z')\vec{r}}{r^3} dS' \} \quad (26)$$

где $\vec{i}(x',y',z') = \vec{i}_M(x',y',z') + \vec{i}_c(x',y',z')$ и $\vec{i}_c(x',y',z')$ – линейная плотность поверхностных полного тока и тока проводимости, соответственно,

$$\vec{i}_M = \vec{n} \times \Delta\vec{M}_\tau \qquad (27)$$

– линейная плотность поверхностного тока намагничивания, зависящая от скачка тангенциальной компоненты намагниченности,

$$\sigma_M(x',y',z') = \vec{n}\,\Delta\vec{M}_n \qquad (28)$$

– поверхностная плотность ФМЗ, определяемая скачком нормальной компоненты намагниченности $\vec{M}$ при переходе из пространства $\Omega$ наружу через поверхность $\Sigma$ (как и раньше, $\vec{n}$ – единичный вектор внешней нормали к $\Sigma$). Отметим, что в формулах (21), (27), (28) можно записывать полные вектора $\vec{P}$ и $\vec{M}$ вместо их нормальных или тангенциальных компонент, но в методических целях мы предпочитаем явно указывать существенную компоненту вектора.

Формулы (15), (20), (25), (26) являются достаточными, чтобы по заданным (или вычисляемым в процессе решения) источникам (зарядам и токам) найти поля $\vec{E}$, $\vec{D}$, $\vec{B}$, $\vec{H}$.

Как показывает формула (20), источниками поля $\vec{D}$ в общем случае являются свободный заряд $\rho_f$, $\sigma_f$ и ФПТ $\vec{j}_P$, $\vec{i}_P$ (см. (17) и (21)). Поэтому утверждения (У1а), (У1б) (см. раздел "Введение") будут справедливыми, только если ФПТ отсутствует. Отметим, что в [3] есть примечание мелким шрифтом: "...мы не станем здесь рассматривать ... явлений пиро- и пьезоэлектричества" (см. [3], с.104)). Но даже это примечание не спасает (У1б) от критики, так как, например, в задаче о шаре из линейного изотропного диэлектрика в однородном внешнем поле нельзя объяснить появление внутри шара поля $\vec{D}$ свободными зарядами. Тем более, что линии $\vec{D}$ в данном случае замкнуты.

В магнитостатике, согласно формуле (26), источниками поля $\vec{H}$ в общем случае являются ток проводимости $\vec{j}_c$, $\vec{i}_c$ и ФМЗ $\rho_M$, $\sigma_M$. Полезность введения ФМЗ в теорию становится особенно очевидной при описании систем с постоянными магнитами.

## 4. Классификация источников электростатического и магнитостатического полей

Как следует из формулы (15) единственным источником напряжённости электростатического поля $\vec{E}$ является электрический заряд. В макроскопической теории электромагнетизма электрические заряды принято подразделять на поляризационные, обусловленные явлением поляризации диэлектриков и математически выражаемые через вектор поляризации (см. (7)), и все прочие заряды, которым следовало бы дать общее название "неполяризационные". Однако в современной учебной и научной литературе заряды первого типа чаще всего называют "связанными", а заряды второго типа - "свободными". В русскоязычной литературе начало этой традиции положил, по-видимому, И.Е.Тамм. Уже в первом издании [16] его учебника можно прочесть: "под свободными зарядами мы будем понимать все электрические заряды за исключением только зарядов диполей, индуцируемых в молекулах поляризованного диэлектрика. Заряды же этих диполей, связанных с молекулами диэлектрика мы будем называть зарядами связанными" ([16], с.113). В дальнейшем И.Е.Тамм детализировал определение свободных зарядов: "под свободными зарядами мы будем понимать, во-первых, все электрические заряды, которые под влиянием электрического поля могут перемещаться на макроскопические расстояния (электроны в металлах и вакууме, ионы в газах и электролитах и т.п.), и, во-вторых, заряды, нанесённые извне на поверхность диэлектриков и нарушающих их нейтральность. Сюда же относятся, например, заряды внутриионной решётки твёрдых диэлектриков, образовавшиеся благодаря недостатку в данном участке диэлектрика ионов определённого знака, так что этот участок в целом уже не нейтрален" ([3], с.101). Близкое по смыслу определение "свободных" и "связанных" зарядов приведено в учебниках Д.В. Сивухина ([4], с.57) и И.Е. Иродова ([5], с.64), однако И.Е. Иродов называет "свободные" ("по Тамму") заряды "сторонними", оставляя за "свободными" зарядами свойство перемещаться под действием электрического поля на макроскопическое расстояния. Р. Фейнман даёт такие определения: "Обозначим ...через $\rho_{пол}$ заряды, появляющиеся за счёт неоднородной поляризации, а остальную часть назовём $\rho_{своб}$. Обычно $\rho_{своб}$ означает заряд, сообщаемый проводникам или распределённый известным образом в пространстве" ([17], с.208). Таким образом, Р. Фейнман относит к "свободным" все заряды, не дающие вклад в правую часть (7). При этом он не ограничивает принадлежность свободных зарядов исключительно проводникам.

Сравнение определений "свободных" и "связанных" зарядов в различных учебниках показывает, что буквальный смысл используемого прилагательного не всегда адекватно описывает свойства объекта. Определение "свободных" зарядов "по Тамму" включает в их число заряды, неспособные перемещаться на макроскопические расстояния, что плохо соотносится с понятием "свободы". Определение "сторонних" зарядов "по Иродову" включает в их состав и собственные (а не привнесённые извне, как можно ожидать из названия) заряды проводника. Так, заряды, возникающие, на поверхности проводника в явлении электростатической индукции в результате действия внешнего электрического поля, следует отнести к "сторонним". Этот пример показывает также, что нарушение электронейтральности тела как целого также не может служить однозначным критерием отнесения зарядов к "свободным" ("по Тамму") и отличать их от "поляризационных", для которых условие сохранения электронейтральности (поляризованного тела) всегда выполняется. Термин "связанные" заряды также не вполне специфичен, так как и "свободные" ("по Тамму") заряды могут локализованы ("связаны") в каком-то месте поверхности или объема тела. Такими будут, например, заряды на поверхности диэлектрика, возникшие в результате контактной электризации (чаще называемой электризацией трением). Возможно, поэтому Д.В. Сивухин и Р. Фейнман используют термин "поляризационные заряды" как синоним к термину "связанные заряды". Ещё один вариант названия предлагает Э. Парселл: "заряды, являющиеся неотъемлемыми частями атомов или молекул диэлектрика, называют обычно "связанными" зарядами. Лучше было бы назвать их структурными зарядами. Эти заряды не могут перемещаться, но являются более или менее упруго связанными и вносят благодаря испытываемому ими смещению вклад в поляризацию" ([18], с.311).

Отмеченное отсутствие единообразия в терминах не играет особой роли, если в каждой конкретной задаче в уравнения (15) и (20) подставляются "правильные" заряды, независимо от их названия. Интересно, что И.Е. Тамм в издании 1929 года [16] указывает, что вводимая им классификация зарядов отличается от принятой в то время. Так, согласно книге Абрагама и Беккера, существуют "истинные" и "свободные" заряды ([19], с.81). В современной классификации ("по Тамму") им соответствуют "свободный" и полный ("свободный"+"связанный") заряды, соответственно. Далее в тексте, как правило, будет использоваться классификация зарядов, предложенная И.Е. Таммом.

В следующих разделах будет показано, что некорректность утверждений (У2а)-(У2в) обусловлена использованием термина "свободный заряд" (или "сторонний заряд") в качестве синонима заряда на проводниках, т.е. понимания термина "свободный заряд" в ограниченном смысле. Если свободный заряд локализован на поверхности диэлектриков, эти утверждения, вообще говоря, неверны, так как изменение напряжённости при заполнении пространства диэлектриком зависит от формы и взаимного расположения заряженных тел. Недостатков утверждений (У2а)-(У2в) лишена формулировка Э. Парселла: "Относительно проводников любой формы, расположенных как угодно и полностью погружённых в однородный изотропный диэлектрик...можно высказать общее утверждение. Оно заключается в том, что при любых зарядах $Q_1$, $Q_2$, и т.д. на различных проводниках, макроскопическое поле среды $\vec{E}_{среды}$ всюду в диэлектрической среде равно 1/ε, умноженной на $\vec{E}_{вакуума}$, которое существовало бы в вакууме с теми же зарядами на тех же проводниках" ([18], p.304).

Для классификации токов в магнитостатике характерна большая схожесть определений, используемых разными авторами, по сравнению с электростатикой. Так, согласно И.Е. Тамму ([3], с.278), И.Е. Иродову ([5], с.168) и Д.В. Сивухину ([4], с.243) полный ток, задающий индукцию магнитного поля соотношением (25), равен сумме тока проводимости и тока намагничивания, определяемого с помощью (10). Словесное описание токов проводимости также сходно у разных авторов. Так, согласно И.Е. Тамму токи проводимости соответствуют "движению зарядов, переносящих макроскопический ток (свободные электроны в металлах, ионы в электролитах и газах)" ([3], с.278). Определение Д.В. Сивухина токов проводимости почти такое же: "Обычные токи, текущие по проводам, связаны с перемещением в веществе носителей тока - электронов или ионов. Эти токи называются токами проводимости." ([4], с.243). Про токи намагничивания Д.В. Сивухин пишет следующее: "Для вычисления макроскопического поля $\vec{B}$ молекулярные токи можно...как-то сгладить, заменив их макроскопическими токами, плавно изменяющимися в пространстве. Такие макроскопические токи называются токами намагничивания" ([4], с.243). Примерно так же описывает токи намагничивания и токи проводимости И.Е.Иродов ([5], с.164). Р. Фейнман подразделяет ток в магнитостатике на две части:"...$j_{маг}$, которая описывает усреднённые токи внутри намагниченных материалов, и дополнительный член $j_{пров}$...который будет описывать всё остальное. Он, вообще говоря, относится к токам в проводниках, но может описывать и другие токи, например, токи зарядов, движущихся свободно через пустое пространство" ([20], с.136).

Таким образом, ток намагничивания можно задать однозначно соотношением (10). Ток проводимости $\vec{j}_c$ можно определить либо методом исключения - макроскопический ток, не описываемый уравнением (10), либо перечислением всех возможных вариантов. Второй подход не гарантирует полноту. Например, в определениях, приведённых выше, не упоминался макроскопический ток, обусловленный замкнутым движением свободных зарядов, полученных электризацией трения, или "замороженных" поляризационных зарядов на поверхности вращающегося диэлектрического тела (см., например, [2], задачи 6.11, 6.21, с.285). Такой ток, безусловно, следует отнести к токам проводимости, так как он не связан с намагниченностью тела.

Ограниченность применимости утверждений (У3а), (У3б) обусловлена, так же как в случае (У2а)-(У2в), "терминологической" причиной, но другого рода, а именно, отсутствием чёткого разграничения между понятием "ток" и "линейный ток". Как будет показано ниже, если ток проводимости течёт в проводах ненулевой толщины, необходимо, в общем случае, учитывать возникающий на их поверхности (границе раздела с линейным изотропным магнетиком) ФМЗ.

Терминологическая неоднозначность при описании источников полей часто сопровождается неоднозначностью их обозначений. Так, во многих учебниках по электромагнетизму ρ обозначает как плотность суммарного (полного) электрического заряда, так и плотность свободного заряда (см. [4], (7.3) и (13.1)]; [5], (1.20) и (3.19), [2], (3.20) и (17.30); [3], (6.5) и (21.5)), а $\vec{j}$ обозначает как плотность суммарного (полного) электрического тока, так и плотность тока проводимости (см. [4], (56.1) и (81.4); [5], (6.26) и (7.13); [2], (35.14) и (38.22а); [3], (47.3) и (65.7)). По-видимому, правильнее поступают авторы, не допускающие (особенно в учебной литературе, где это важно методически) такого дублирования обозначений различных физических величин и использующие "говорящие" подстрочные индексы (см. [18], (59), с.313 и (53), с.365; [17], (10.18); [20], (36.15)].

Заканчивая обсуждение особенностей терминологии, отметим, что векторам $\vec{D}$ и $\vec{H}$ иногда отказывают в праве называться "полем", так как они являются комбинацией (см. следующий раздел) "настоящих" (или "первичных") полей $\vec{E}$ и $\vec{B}$, описывающих силовое воздействие на заряды и токи (см., однако, дискуссию в [21]), и векторов $\vec{P}$ и $\vec{M}$, характеризующих состояние вещества ([2], с.144, с.269). При этом, с математической точки зрения $\vec{D}$ и $\vec{H}$, несомненно, являются векторными полями (векторными функциями, заданными в трехмерном пространстве).

## 5. Свободные и связанные заряды в задачах электростатики

Обычно при решении задачи о нахождения полей $\vec{E}$, $\vec{D}$, $\vec{B}$, $\vec{H}$ известны только некоторых из их источников. В электростатических задачах с проводниками (в вакууме или в линейном изотропном диэлектрике), равновесное распределение свободных зарядов на поверхности проводников можно найти, решая уравнение Лапласа с заданными граничными условиями. Иногда, используя соображения симметрии, можно угадать правильное распределение свободных зарядов на поверхности проводников, а затем подтвердить его, проверив выполнение формул (3), (6) и условий сохранения суммарных зарядов проводников и эквипотенциальности. Теорема единственности ([22], с.90) гарантирует отсутствие другого решения. Если свободные заряды расположены на поверхности диэлектрика, то они считаются заданными благодаря известной процедуре электризации. Заданными считаются обычно ФМЗ постоянных магнитов и связанный заряд, обусловленный "замороженной" электрической поляризацией диэлектриков (электретов, пиро- и сегнетоэлектриков). В магнитостатике, кроме того, обычно считают заданным ток проводимости.

Индуцированный заряд (связанный в диэлектриках и ФМЗ в магнетиках), а также ток намагничивания и ФПТ, как правило, заранее неизвестны и их можно найти только в процессе решения задачи. В ряде случаев можно получить простые соотношения, устанавливающие связи между различными типами источников полей. Это, в частности, относится к системам, включающим линейный изотопный диэлектрик, для которого постулируется выполнение соотношений:

$\vec{D} = \varepsilon\varepsilon_0 \vec{E}$ \hfill (29)

$\vec{P} = (\varepsilon-1)\varepsilon_0 \vec{E} = \chi_E \varepsilon_0 \vec{E}$ \hfill (30)

где ε - диэлектрическая проницаемость, $\chi_E = \varepsilon - 1$ - диэлектрическая восприимчивость, или линейный изотропный магнетик, для которого выполняется:

$$\vec{B} = \mu\mu_0 \vec{H} \qquad (31)$$

$$\vec{M} = (\mu - 1)\vec{H} = \chi_H \vec{H} \qquad (32)$$

где μ - магнитная проницаемость, $\chi_H = \mu - 1$ - магнитная восприимчивость.

Далее, для краткости, будем называть среды, подчиняющиеся условиям (29)-(32), простыми: диэлектриком или магнетиком, соответственно (см., например, [23], с.45). Кроме того, будем считать одну из сред жидкой (газообразной), чтобы избежать необходимости уточнения особенностей контакта двух твердых тел. Обычно твердое тело (проводник, диэлектрик, магнетик) находится внутри жидкой среды.

Отметим, что система уравнений (29), (30) удовлетворяет соотношению:

$$\vec{D} = \varepsilon_0 \vec{E} + \vec{P} \qquad (33)$$

которое выполняется для любых диэлектриков, так как служит определением вектора $\vec{D}$.

Аналогично, система (31), (32) удовлетворяет соотношению:

$$\vec{B} = \mu_0(\vec{H} + \vec{M}) \qquad (34)$$

которое выполняется для любых магнетиков, так как служит определением вектора $\vec{H}$.

Соотношения (33) и (34) предполагают, что мы пренебрегаем при описании диэлектрика и магнетика моментами с порядком выше дипольного (квадрупольными и т.д.) [24].

Для простых однородных диэлектриков из (6), (7), (29), (30) при условии ε=const можно получить:

$$\rho_b = -\rho_f(\varepsilon - 1)/\varepsilon \qquad (35).$$

Из (35) следует,

$$\rho = \rho_b + \rho_f = \rho_f/\varepsilon \qquad (36)$$

т.е. в простом однородном диэлектрике объемная плотность полного заряда меньше объёмной плотности свободного заряда в ε раз.

Строго говоря, соотношения (35) и (36) относятся к точечному свободному заряду. Их можно интерпретировать так: если равномерно заряженное (свободным зарядом) малое тело объёмом *V* помещается в диэлектрик, то на его границе возникают связанные заряды (35), которые совместно со свободным зарядом тела обеспечивают среднюю плотность (36) полного заряда в объёме *V*. При *V*→0 форма тела и распределение свободного заряда по объёму *V* не играют существенного значения и средняя плотность становится плотностью в "точке". Разумеется, "стремление к нулю" объёма следует понимать в смысле "физически малого объёма" ([3], с.19).

Возможен ещё один аргумент в пользу формулы (35): величина индуцированных связанных зарядов в однородном простом диэлектрике вблизи точечного свободного заряда зависит только от локальной плотности этого заряда (и не зависит от остальных зарядов в системе), так как модуль напряжённости поля малого заряженного тела (моделью которого служит точечный заряд) неограниченно возрастает при приближении к нему (как $\sim 1/r^2$) и поэтому поля остальных (удалённых) зарядов можно не учитывать.

Если проводник, погружённый в простой жидкий диэлектрик, имеет конечные размеры (и уже не может рассматриваться как точечный заряд), соотношения аналогичные (35) и (36) выполняются для поверхностных плотностей свободного заряда на проводнике и связанного заряда в диэлектрике в месте, где он вплотную примыкает к поверхности проводника:

$\sigma_b = -\sigma_f((\varepsilon-1))/\varepsilon$ (37)

$\sigma = \sigma_b + \sigma_f = \sigma_f/\varepsilon$ (38)

Формулы (37) и (38) справедливы не только для однородного, но и для неоднородного простого диэлектрика. Они могут быть получены из граничных условий, которые, в свою очередь, являются результатом применения интегральной теоремы Гаусса к (6) и (7) и имеют вид:

$P_{2n} - P_{1n} = -\sigma_b$ (39)

$D_{2n} - D_{1n} = \sigma_f$ (40)

Единичный вектор нормали $\vec{n}$ в граничных условиях (39) и (40) (и в других аналогичных формулах) направлен от среды 1 к среде 2. Для получения (37), (38) из (39), (40), нужно воспользоваться связью векторов $\vec{D}$ и $\vec{P}$ для простого диэлектрика (см. (29) и (30)):

$\vec{P} = \vec{D}(\varepsilon - 1)/\varepsilon$ (41)

и условие отсутствия внутри проводника электрического поля, приводящего к

$D_{1n} = 0; P_{1n} = 0$ (42).

Если свободный заряд расположен на поверхности непроводящего тела, то соотношения (42), а вместе с ним и формулы (37), (38), в общем случае перестают выполняться. Исключение представляют некоторые системы с симметричным расположения заряда. Например, изолированная равномерно заряженная по поверхности непроводящая сфера создаёт вне и внутри себя такое же электрическое поле, как и проводящая.

Если граница разделяет простые диэлектрики с проницаемостью $\varepsilon_1$ и $\varepsilon_2$, соответственно, то граничное условие (40) для нормальной компоненты вектора $\vec{D}$ можно, с учётом (29), записать в виде:

$\varepsilon_2 \varepsilon_0 E_{n2} - \varepsilon_1 \varepsilon_0 E_{n1} = \sigma_f$ (43)

или

$$\varepsilon_2\varepsilon_0(E_n^{вн}+(\sigma_b+\sigma_f)/(2\varepsilon_0))-\varepsilon_1\varepsilon_0(E_n^{вн}-(\sigma_b+\sigma_f)/(2\varepsilon_0))=\sigma_f \quad (44)$$

где $E_{n1}$ и $E_{n2}$ – проекция на нормаль к границе напряженности электрического поля в первой и второй среде, соответственно (в точках, бесконечно близких к границе),

$$E_n^{вн}=(E_{n1}+E_{n2})/2 \quad (45)$$

- нормальная компонента напряженности внешнего поля в точке непосредственно на границе. Под "внешним" понимается поле, создаваемое всеми зарядами, кроме свободных $\sigma_f$ и связанных $\sigma_b$ зарядов на малой площадке в рассматриваемой точке границы (граница подразумевается достаточно гладкой).

Из (44) следует формула для плотности связанного заряда:

$$\sigma_b=\sigma_f[2/(\varepsilon_1+\varepsilon_2)-1]+2\varepsilon_0 E_n^{вн}(\varepsilon_1-\varepsilon_2)/(\varepsilon_1+\varepsilon_2) \quad (46)$$

или для полного заряда:

$$\sigma=\sigma_b+\sigma_f=2\sigma_f/(\varepsilon_1+\varepsilon_2)+2\varepsilon_0 E_n^{вн}(\varepsilon_1-\varepsilon_2)/(\varepsilon_1+\varepsilon_2) \quad (47)$$

Так как величина $E_n^{вн}$ выражается через интеграл (15), уравнение (47) задаёт весьма сложную в общем случае систему интегральных уравнений ([1], с.258).

Для симметричного расположения зарядов на поверхности тела граничные условия (43), (44) могут упроститься вследствие $E_{n1}=0$. Например, для сферы, поверхность которого равномерно покрывает свободный заряд, с учётом $E_{n2}=(\sigma_b+\sigma_f)/\varepsilon_0$; $E_n^{вн}=(\sigma_b+\sigma_f)/2\varepsilon_0$, уравнение (44) преобразуется в (38), в котором $\varepsilon$ следует заменить на $\varepsilon_2$. При этом материал сферы (проводник или диэлектрик) неважен.

Можно привести и другие примеры упрощения (43), (44). Например, пусть свободный заряд равномерно распределён с плотностью $\sigma_f$ по торцу стержня из твёрдого простого диэлектрика с проницаемостью $\varepsilon_1$, расположенного в простом однородном жидком диэлектрике с проницаемостью $\varepsilon_2$, занимающем всё пространство вне стержня (рис.1).

Если стержень длинный ($L>>R$), то для нахождения напряжённости поля в точке А около центра торца, на малом ($R>>h$) расстоянии $h$ от поверхности, можно воспользоваться моделью "бесконечной заряженной плоскости", так как вкладами в напряжённость поля от индуцированных связанных зарядов $\sigma_b'$ и $\sigma_b''$ на других участках границы можно пренебречь. Заряды на правом плоском торце также не вносят вклад в $E_n^{вн}$, поскольку их поля при $h\to 0$ направлены по касательной к границе в рассматриваемой точке. Пренебрегая в (44) $E_n^{вн}$ по сравнению с $(\sigma_b+\sigma_f)/2\varepsilon_0$, получаем:

$$\sigma=2\sigma_f/(\varepsilon_1+\varepsilon_2) \quad (48)$$

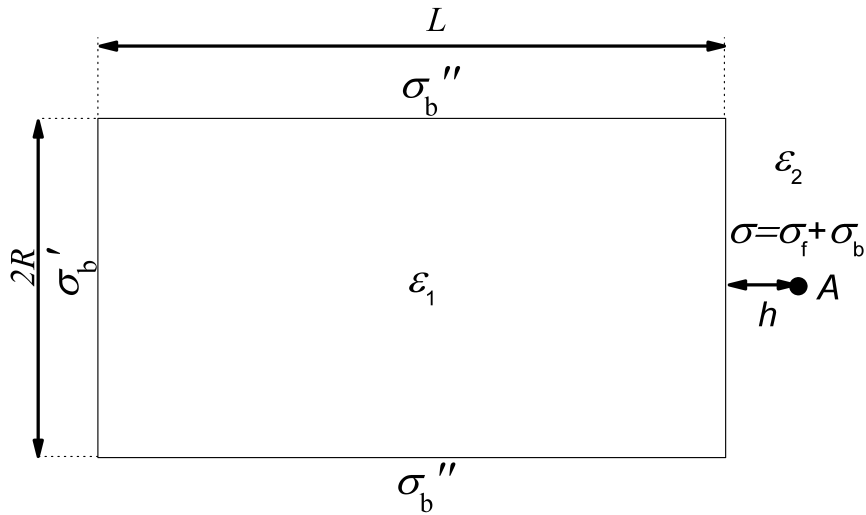

Рис.1 Диэлектрический (проницаемость $\varepsilon_1$) цилиндрический стержень, на правом торце которого равномерно распределён свободный заряд с плотностью $\sigma_f$. На границах стержня с окружающим простым диэлектриком с проницаемостью $\varepsilon_2$, возникают связанные заряды $\sigma_b$, $\sigma_b'$, $\sigma_b''$. Электрическое поле ищется в точке A, расположенной на малом расстоянии $h \ll R$ от центра правого торца. Радиус стрежня $R$, длина $L$. В статье рассматриваются два предельных случая: $L \gg R$ (модель "длинного стержня") и $R \gg L$ (модель "тонкого диска").

Напряжённость поля в точке A равна:

$$E = \sigma/(2\varepsilon_0) = \sigma_f/\varepsilon_0(\varepsilon_1 + \varepsilon_2) \qquad (49)$$

По сравнением с вакуумом ($\varepsilon_2 = 1$) напряжённость поля уменьшилась в $(\varepsilon_1 + \varepsilon_2)/(\varepsilon_1 + 1)$ раз. Если $\varepsilon_1 \approx 1$ и $\varepsilon_2 = \varepsilon \gg 1$, то $(\varepsilon_1 + \varepsilon_2)/(\varepsilon_1 + 1) \approx \varepsilon/2$.

Этот пример иллюстрирует ограниченность области применимости утверждений (У2а), (У2б), (У2в), показывая, что напряжённость поля свободных зарядов не всегда уменьшается в $\varepsilon$ раз, если эти заряды перенести, сохранив их начальное взаимное расположение, из вакуума в однородный простой диэлектрик с проницаемостью $\varepsilon$, заполняющий всё пространство (кроме объёма тех тел, которым принадлежат свободные заряды).

Если для стержня на рис.1 выполняется $L \ll R$, т.е. он представляет собой тонкий диск, необходимо учесть вклад в $E_n^{\text{вн}}$ зарядов $\sigma_b'$. Для правого торца на рис.1 уравнение (47) примет вид:

$$\sigma = \sigma_b + \sigma_f = 2\sigma_f/(\varepsilon_1 + \varepsilon_2) + \sigma_b'(\varepsilon_1 - \varepsilon_2)/(\varepsilon_1 + \varepsilon_2) \qquad (50)$$

а для левого торца, на котором нет свободных зарядов

$$\sigma_b' = \sigma(\varepsilon_1 - \varepsilon_2)/(\varepsilon_1 + \varepsilon_2) \qquad (51)$$

Подставляя (51) в (50), получаем:

$$\sigma = \sigma_f(\varepsilon_1 + \varepsilon_2)/(2\varepsilon_1\varepsilon_2) \qquad (52)$$

Напряжённость поля в точке $A$ равна:

$E = (\sigma + \sigma_b')/(2\varepsilon_0) = \sigma[1+(\varepsilon_1-\varepsilon_2)/(\varepsilon_1+\varepsilon_2)]/(2\varepsilon_0) = \sigma_f(\varepsilon_1+\varepsilon_2)[1+(\varepsilon_1-\varepsilon_2)/(\varepsilon_1+\varepsilon_2)]/(4\varepsilon_0\varepsilon_1\varepsilon_2) =$

$= \sigma_f/(2\varepsilon_0\varepsilon_2)$ \hfill (53)

Формула (53) интуитивно понятна: напряжённость поля связанных зарядов диэлектрика тонкого диска в рассматриваемой модели "плоского конденсатора" равна нулю вне диска, поэтому связанные заряды в диэлектрике с проницаемостью $\varepsilon_2$ индуцируются только полем свободных зарядов $\sigma_f$. Поэтому в формулу (51) проницаемость $\varepsilon_1$ диэлектрика диска не вошла. Поле вне диска не изменится, если считать диэлектрический тонкий диск проводящим (что эквивалентно $\varepsilon_1 \to \infty$), разумеется, при неизменном положении зарядов $\sigma_f$.

Приведённые примеры демонстрирует зависимость степени уменьшения напряжённости электрического поля свободных зарядов, локализованных на поверхности диэлектрического тела, при помещении его в "бесконечную" диэлектрическую среду, от формы этого тела. Для свободных зарядах на проводниках такой зависимости от формы нет, так как (35) и (36) выполняются независимо от формы проводника.

Покажем теперь, что для диэлектрического тела с "замороженной" поляризацией его влияние на поляризацию окружающего простого диэлектрика также, в общем случае, зависит от формы тела. Пусть среда 2 - жидкий простой диэлектрик с проницаемостью $\varepsilon$, а среда 1 представляет собой твёрдый диэлектрик с "замороженной" поляризаций (электрический аналог постоянного магнита). Граничное условие для нормальной компоненты вектора $\vec{D}$ принимает вид:

$\varepsilon\varepsilon_0 E_{n2} - D_{n1} = \sigma_f$ \hfill (54)

или

$\varepsilon\varepsilon_0(E_{n2}^{\text{вн}} + (\sigma_b + \sigma_f)/(2\varepsilon_0)) - (D_{n1}^{\text{вн}} - \sigma_f/2) = \sigma_f$ \hfill (55)

где $E_{n2}$ и $D_{n1}$ − проекция на нормаль к границе напряженности электрического поля и электрической индукции, соответственно во второй и первой среде (в точках, бесконечно близких к границе), $E_{n2}^{\text{вн}}$ и $D_{n1}^{\text{вн}}$ соответствующие внешние вклады в нормальные компоненты напряженности и электрической индукции, соответственно. Смысл прилагательного "внешний" здесь тот же, что и при обсуждении формулы (44). В формулах (54), (55) учтено, что возможный ФПТ, локализованный на рассматриваемом малом участке поверхности, не даёт вклад в поток при выводе граничного условия для нормальной компоненты вектора $\vec{D}$. Величина $E_{n2}^{\text{вн}}$ зависит от всех внешних зарядов (свободных и связанных), а $D_{n1}^{\text{вн}}$ только от внешних свободных зарядов и от внешних фиктивных поляризационных токов (см. (15) и (19), соответственно). Отметим, что величину $D_{n1}^{\text{вн}}$ можно, при необходимости, выразить через нормальные проекции напряжённости внешнего поля $E_{n1}^{вн}$ и поляризации $P_{n1}$ в первой среде. Вектор поляризации обычно считается известным в таких задачах. При этом можно использовать связь между проекциями векторов в (33) на нормаль к границе: $D_{n1}=\varepsilon_0 E_{n1}+P_{n1}$, откуда следует

$$D_{n1}^{вн}=\varepsilon_0 E_{n1}^{вн}-\sigma_b/2+P_{n1} \qquad (56)$$

Рассмотрим два частных случая, когда уравнения (54), (55) упрощаются. Применим уравнение (55) к диэлектрическому стержню (рис.2), в котором вектор $\vec{P}$ =const перпендикулярен торцам, а свободные заряды на поверхности отсутствуют ($\sigma_f=0$).

Рассмотрим сначала случай $L>>R$ (модель "длинного стержня"). Основываясь на рассуждениях, приведших к (48) и (49), можно пренебречь $E_{n2}^{вн}$ в (55) по сравнению с $\sigma_b/(2\varepsilon_0)$ и $\varepsilon_0 E_{n1}^{вн}$ в (56) по сравнению с $P_{n1}$ и $\sigma_b/2$. Подставляя в (54) $D_{n1}^{вн}$ из (56), получаем формулу для плотности связанного заряда на границе стержня с простым диэлектриком с проницаемостью ε вблизи точки A (рис.2):

$$\sigma_b=2P/(1+\varepsilon) \qquad (57)$$

Напряжённость поля в точке A равна

$$E=E_{2n}^{вн}+\sigma_b/(2\varepsilon_0)=\sigma_b/(2\varepsilon_0)=P/\varepsilon_0(1+\varepsilon) \qquad (58)$$

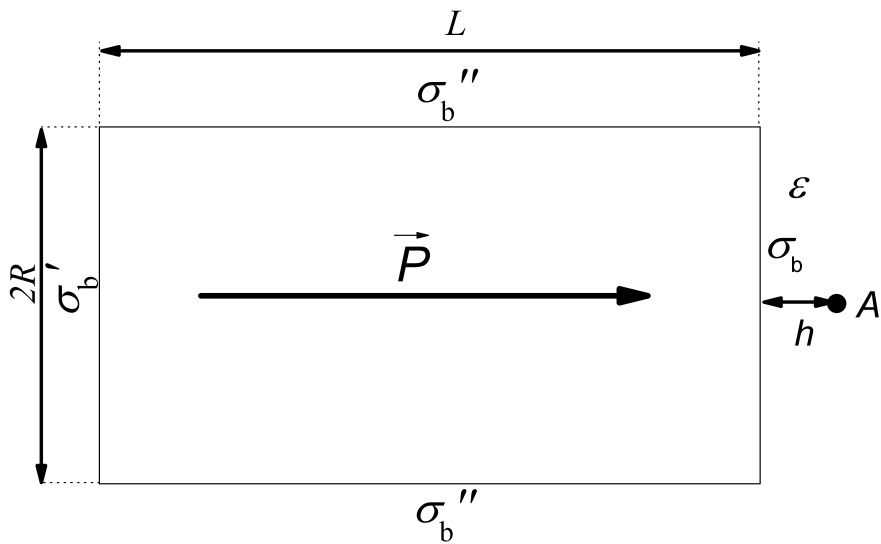

Рис.2 Диэлектрический цилиндрический стержень с однородной "замороженной" поляризацией (вектор поляризации $\vec{P}$). На границах стержня с окружающим простым диэлектриком с проницаемостью ε, возникают связанные заряды $\sigma_b$, $\sigma_b'$, $\sigma_b''$. Электрическое поле ищется в точке A, расположенной на малом расстоянии $h<<R$ от центра правого торца. Радиус стрежня $R$, длина $L$. В статье рассматриваются два предельных случая: $L>>R$ (модель "длинного стержня") и $R>>L$ (модель "тонкого диска").

Как следует из (58), напряжённость в точке А уменьшилась в (ε+1)/2 раз в сравнении с её значением для стержня в вакууме. Индукция в торце стержня $D=D_{n1}=D_{n2}$ равна, согласно (58) и (29):

$D=\varepsilon P/(1+\varepsilon)$  (59)

Если бы стержень находился в вакууме, величину индукции $D=P/2$, которая получается из (59) при ε=1, можно было бы интерпретировать как поле (20) протекающего по поверхности стержня кругового замкнутого ФПТ с линейной плотностью $i_P=P$ (см. (21)). Такая система эквивалентна "полубесконечному" тонкому соленоиду с плотностью поверхностного тока проводимости $nI=P$ (n − плотность витков в обмотке, I − сила тока в тонком проводе, формирующем обмотку). Если бесконечный соленоид представить в виде последовательного сочленения (без зазора) двух одинаковых "полубесконечных" соленоидов, то, применяя принцип суперпозиции, можно заключить, что напряжённость магнитного поля в торце "полубесконечного" соленоида равна половине напряжённости поля внутри бесконечного соленоида, т.е. $nI/2$. Такие же рассуждения применимы для рассматриваемого поляризованного стержня.

Так как ε/(1+ε)>1/2, формула (59) указывает, что размещение стержня с продольной "замороженной" поляризацией в простом безграничном диэлектрике вызывает появление на боковой границе стержня дополнительного ФПТ $i_P'$, сонаправленного с $i_P$ и поэтому усиливающего $\vec{D}$ в точке А. Дополнительный ток $i_P'$ обусловлен возникшей в внешней среде поляризацией (см. (21)). Причиной этой поляризации являются, в конечном итоге, реальные связанные заряды на торцах стержня. Поэтому величина $i_P'$ должна уменьшаться при удалении от торцов (минимальна в середине стержня). В модели "бесконечного" продольно поляризованного цилиндра с $\vec{P}$=const реальные электрические заряды удалены в "бесконечность", электрическое поле $\vec{E}$ вне цилиндра отсутствует (как отсутствует магнитное поле $\vec{B}$ вне идеального бесконечного соленоида), состояние диэлектрика вне цилиндра не изменяется, поэтому $i_P'$=0, электрическая индукция внутри цилиндра определяется только током $i_P$ и равна $\vec{D}=\vec{P}$=const.

Отметим, что формулу (59) можно получить из формулы (49) для напряжённости, подставляя в неё $\varepsilon_1$=1, $\varepsilon_2$=ε, $\sigma_f$=P и умножив, согласно (29), на $\varepsilon\varepsilon_0$. Другими словами, возникающий на торце стержня благодаря "замороженной" поляризации связанный заряд, равный P для стержня в вакууме, порождает в диэлектрике такое же поле, как свободный заряд той же величины, если диэлектрическую восприимчивость материала стержня принять равной единице. Аналогичное свойство характерно и для постоянных магнитов ([25], с.253).

Для однородно поляризованного тонкого диска ($L<<R$ на рис.2), можно считать, что нормальная компонента электрической индукции $D_{1n} \approx 0$, так как создающий её ФПТ расположен на ребре диска с очень малой площадью, находясь при этом на большом расстоянии от центра основания диска ($R>>h$). Поэтому из (54) следует $E_{n2} \approx 0$, т.е. электрическое поле снаружи диска отсутствует. Этот результат, как и (53), интуитивно понятен - связанные заряды, обусловленные скачком нормальной компоненты вектора $\vec{P}$ "замороженной" поляризации на основаниях диска, образуют систему, подобную плоскому конденсатору, электрическое поле которой вне конденсатора пренебрежимо мало.

В примерах, включающих диэлектрики с "замороженной" поляризацией, при обсуждении источников вектора электрической индукции $\vec{D}$ мы использовали понятие ФПТ. В следующем разделе оба вида источников поля $\vec{D}$ будут рассмотрены более подробно.

## 6. Два вида источников поля $\vec{D}$: свободные заряды и "фиктивные поляризационные токи"

Вектор электрической индукции $\vec{D}$ (другое название - вектор электрического смещения) часто рассматривается как "вспомогательный вектор, не имеющий глубокого физического смысла" ([5], с.70), главное достоинство которого заключается в том, что его поток через любую замкнутую поверхность пропорционален свободному заряду в объёме, ограниченном этой поверхностью ([4], с.60). Действительно, в системе зарядов, обладающей сферической, цилиндрической или плоской симметрией, вектор $\vec{D}$ не зависит от связанных зарядов и его легко найти, применив интегральную форму теоремы Остроградского-Гаусса. Например, размещение простого однородного диэлектрика между обкладками конденсатора (плоского, сферического или цилиндрического) оставляет поле $\vec{D}$ таким же, каким оно было в отсутствии диэлектрика (для плоского и цилиндрического конденсаторов это справедливо лишь в пренебрежении краевыми эффектами). Некоторые авторы учебников делают замечание, что $\vec{D}$ зависит в общем случае от поляризационных зарядов, не уточняя явный вид этой зависимости.

Как следует из (20), источниками поля $\vec{D}$ являются свободный заряд и ФПТ: объёмный (17) или поверхностный (21) ([8], [26]). Поэтому утверждения (У1а) и (У1б) без указания области их применимости могут ввести в заблуждение.

Пример электрической системы, для которой поле $\vec{D}$ зависит только от ФПТ - тороидальный диэлектрик с "замороженной" поляризацией, в котором линии $\vec{P}$ направлены по окружностям (рис.3).

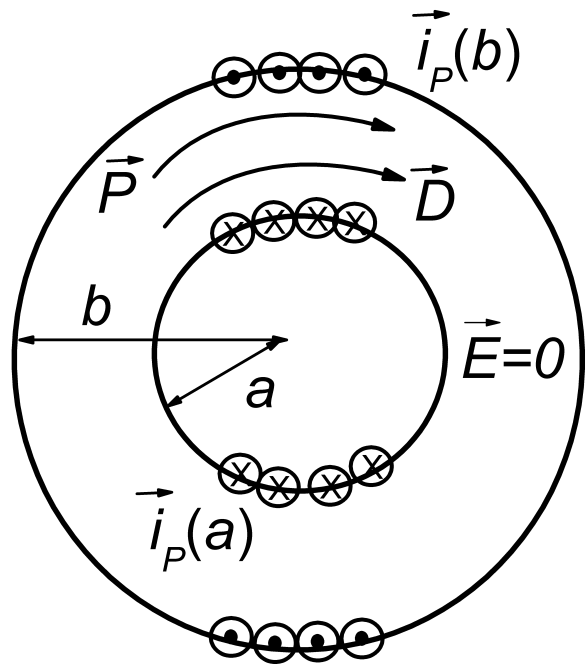

Рис.3 Диэлектрическое тело в форме тороида с "замороженной" поляризацией $\vec{P}$. Линии $\vec{P}$ представляют собой окружности. Кружки показывают направление поверхностного ФПТ с линейной плотностью $i_P(r)=P(r)\sim 1/r$, где $r$ – расстояние до центра тороида. Внутри тороида: $\vec{E}=0$, $\vec{D}=\vec{P}$. Вне тороида: $\vec{E}=0$, $\vec{D}=0$, $\vec{P}=0$.

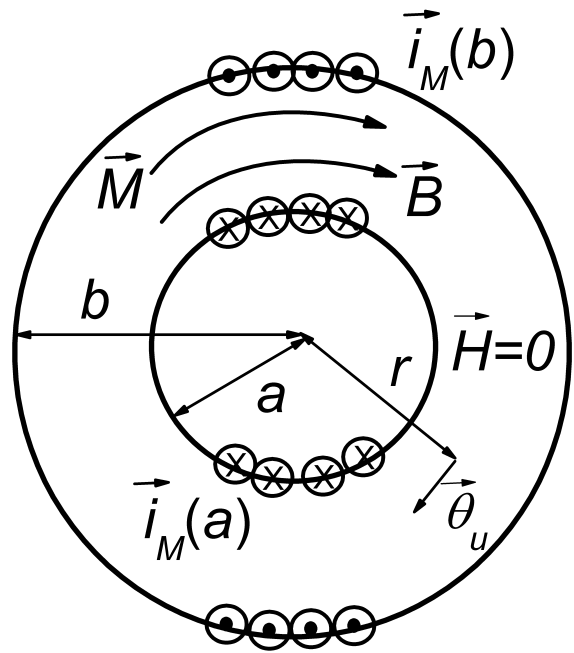

Рис.4 Тороидальный постоянный магнит. Вектор $\vec{M}$ направлен вдоль $\vec{\theta}_u$, где $\vec{\theta}_u$ – единичный вектор касательный к окружности с центром на оси симметрии тороида. Кружки показывают направление поверхностного тока намагничивания $i_M(r)=M(r)\sim 1/r$, где $r$ – расстояние до центра тороида. Внутри тороида: $\vec{H}=0$, $\vec{B}=\mu_0\vec{M}$. Вне тороида: $\vec{B}=0$, $\vec{H}=0$, $\vec{M}=0$.

Нормальные к поверхности диэлектрика компоненты $\vec{P}$ равны нулю. Это электростатический аналог тороидального постоянного магнита (рис.4). Его можно создать, например, "нанизав" сегнетоэлектрический тороид на длинный круглый соленоид с изменяющимся током, создающим круговое вихревое электрическое поле. Это поле и поляризует сегнетоэлектрик нужным образом. В системе на рис.3 напряжённость электростатического поля равна нулю во всех точках пространства (так как нет ни свободных, ни связанных зарядов), при этом внутри тороида $\vec{D}=\vec{P}$ (в магнитном аналоге на рис.4 $\vec{B}=\mu_0\vec{M}$).

Если сделать узкий (ширина $d$ много меньше среднего радиуса тороида) поперечный зазор в тороиде (рис.5), в нём возникнет поле с напряжённостью $\vec{E}_g \approx \vec{P}/\varepsilon_0$ (в магнитном аналоге $\vec{H}_g \approx \vec{M}$, рис.6).

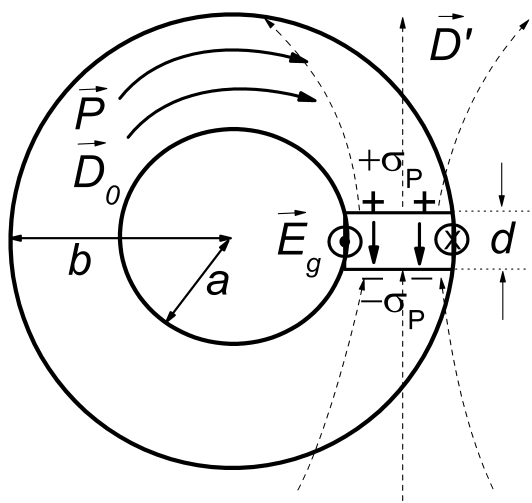

Рис.5 Диэлектрическое тело в форме тороида, такое же как на рис.3, но с узким зазором шириной $d<<(b-a)$. $\vec{E}_g$ — напряжённость электрического поля внутри зазора. Его источник – связанные заряды $\sigma_b$ в местах скачка нормальной компоненты $\vec{P}$. Кружки показывают направление поверхностного ФПТ, создающего поле $\vec{D}'$ (штриховые линии). Поле $\vec{D}_0$ ФПТ, "текущего" по всей на боковой поверхности тороида (включая зазор), сосредоточено внутри тороида также, как на рис.3. Суммарное поле вектора электрической индукции $\vec{D}=\vec{D}_0+\vec{D}'$.

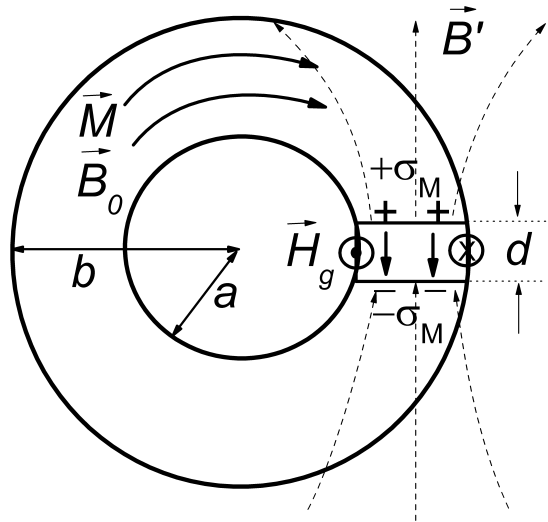

Рис.6 Тороидальный постоянный магнит, такой же как на рис.4, но с узким зазором шириной $d<<(b-a)$. $\vec{H}_g$ – напряжённость магнитного поля внутри зазора. Его источник – ФМЗ $\sigma_M$ в местах скачка нормальной компоненты $\vec{M}$. Кружки показывают направление поверхностного тока намагничивания, создающего поле $\vec{B}'$ (штриховые линии). Поле $\vec{B}_0$ тока намагничивания, текущего по всей на боковой поверхности тороида (включая зазор), сосредоточено внутри тороида так же, как на рис.4. Суммарное поле вектора магнитной индукции $\vec{B} = \vec{B}_0 + \vec{B}'$.

Поле вектора $\vec{D}$ в системе на рис.5 равна сумме поля $\vec{D}_0$ тороида без щели (рис.3) и поля $\vec{D}'$ короткого соленоида ("катушки") длины $d$ и диаметра $b-a>>d$ (рис.5). Создающие эти поля ФПТ тороида и ФПТ "катушки" на боковой поверхности зазора компенсируют друг друга. Так как "катушка" тонкая, сила тока в ней мала, и поэтому внутри материала тороида $|\vec{D}'|<<|\vec{D}_0|=|\vec{P}|$. Напряжённость электрического поля снаружи зазора равна $\vec{E}'=\vec{D}'/\varepsilon_0$, причем эта формула справедлива и внутри материала тороида: $\vec{E} = (\vec{D}-\vec{P})/\varepsilon_0 = (\vec{D}_0+\vec{D}'-\vec{P})/\varepsilon_0 = \vec{D}'/\varepsilon_0$.

Так как в зазоре вакуум, то $\vec{E}_g = \vec{D}/\varepsilon_0 = (\vec{D}_0+\vec{D}')/\varepsilon_0 \approx \vec{D}_0/\varepsilon_0 = \vec{P}/\varepsilon_0$. Поэтому для тонкого зазора $|\vec{E}_g|>>|\vec{E}'|$, т.е. электрическое поле "сосредоточено" в основном внутри зазора. В материале тороида вблизи зазора вектор $\vec{E}'$ направлен противоположно вектору $\vec{P}$ и его можно назвать деполяризующим полем (аналог поля размагничивания в задаче с тороидальным постоянным магнитом). Это поле нельзя считать однородным, так как оно быстро убывает с расстоянием. Этот вопрос мы рассмотрим более подробно в разделе 8, обсуждая магнитостатическую систему на рис.6.

Важно отметить, что достаточно простые формулы для $\vec{E}$ и $\vec{D}$ для тороида с зазором (рис.5) получаются, только если поляризация $\vec{P}$ не зависит от внешних полей ("абсолютно жёсткий" сегнетоэлектрик). В противном случае задача усложняется, в частности, нужно учитывать связанные заряды на боковой поверхности тороида, возникающие благодаря ненулевой нормальной компоненте вектора $\vec{E}$'.

Подмеченные выше аналогии между магнитостатической и электростатической задачами не случайны. Электростатические системы, не включающие свободные заряды, могут быть охарактеризованы вектором $\vec{D}$, который задаётся формулой (16) с ФПТ (17), (21) и $\rho_f=0$, $\sigma_f=0$. Аналогичные магнитостатические системы, не содержащие токов проводимости, можно описать вектором $\vec{B}$, задаваемым формулой (25) с реальными токами намагничивания (10), (27) и $\vec{j}_c=0$, $\vec{i}_c=0$. Математическое сходство пар уравнений (16) и (25), (10) и (17), (21) и (27) в этом случае очевидно.

Альтернативный (и чаще используемый) способ описания электростатических систем, не включающих свободные заряды, состоит в расчете вектора $\vec{E}$ по формуле (20) с реальными связанными зарядами (7), (39) и $\rho_f=0$, $\sigma_f=0$. Аналогичные магнитостатические системы, не включающие токи проводимости, описываются вектором $\vec{H}$, задаваемым формулой (26) с фиктивными магнитными зарядами (24), (28) и $\vec{j}_c=0$, $\vec{i}_c=0$. И опять, при указанных условиях, очевидно математическое сходство пар уравнений (20) и (26), (7) и (24), (39) и (28).

Обсудим, в каких случаях при расчёте $\vec{D}$ можно не учитывать ФПТ. В простом диэлектрике объёмная плотность этих токов равна согласно (17):

$$\vec{j}_P = \mathrm{rot}(\vec{P}) = \mathrm{rot}(\varepsilon_0(\varepsilon-1)\vec{E}) = \varepsilon_0\{\nabla\varepsilon\times\vec{E} + (\varepsilon-1)\mathrm{rot}(\vec{E})\} = \varepsilon_0\nabla\varepsilon\times\vec{E}, \qquad (60)$$

где $\varepsilon$ − диэлектрическая проницаемость диэлектрика и учтена потенциальность вектора $\vec{E}$ (см. (3)). Если простой диэлектрик однороден, то $\nabla\varepsilon=0$ и объёмный поляризационный ток отсутствует. На границе двух однородных простых диэлектриков с проницаемостью $\varepsilon_1$ и $\varepsilon_2$, возникает поверхностный ФПТ с линейной плотностью (см. (21)):

$$\vec{i}_P = (\varepsilon_2 - \varepsilon_1)\vec{n}\times\vec{E}_\tau \qquad (61)$$

где $\vec{E}_\tau$ − тангенциальная компонента напряжённости электрического поля на границе.

Таким образом, ФПТ отсутствует в простом диэлектрике, если он однородный и линии поля перпендикулярны к его границам. Эти условия выполняется, например, для системы заряжённых проводников в безграничном однородном простом диэлектрике. Поверхность проводников эквипотенциальна и линии поля перпендикулярны к ней. Если

ФПТ отсутствует, вектор $\vec{D}$ определяется согласно (20) только свободными зарядами. Если эти заряды не изменяют положение в пространстве, то не изменяется и поле $\vec{D}$. В простом диэлектрике это поле создаёт напряженность в ε раз меньшую по сравнению с вакуумом. В учебниках эту особенность часто описывают так: "Если однородный и изотропный диэлектрик полностью заполняет объём, ограниченный эквипотенциальными поверхностями поля сторонних зарядов, то вектор электрического смещения совпадает с вектором напряжённости поля сторонних зарядов, умноженным на $\varepsilon_0$, и, следовательно, напряжённость поля внутри диэлектрика в ε раз меньше, чем напряжённость поля сторонних зарядов" ([27], с.76). В этом утверждении под "сторонними" понимаются свободные заряды.

Важно отметить ещё один аспект. Обычно в учебниках "молчаливо" предполагается, что распределение свободного заряда на поверхности проводников не изменяется, если всё окружающее их пространство заполнить простым однородным диэлектриком. Желательно обосновать это предположение. Для этого следует воспользоваться теоремой единственности ([22], с.90), утверждающей, в частности, что если заданы полные заряды проводников, то существует единственно возможное распределение плотности зарядов по поверхности проводников, удовлетворяющее их эквипотенциальности и уравнениям электростатики, в частности, уравнению Лапласа (в общем случае – Пуассона). Поэтому достаточно любым методом найти такое распределение, чтобы считать его решением задачи.

Пусть равновесные свободные заряды на поверхности проводников в вакууме создают поле $\vec{E}_0(\vec{r})$ с потенциалом $\varphi_0$. В пространстве между проводниками выполняется $\Delta\varphi_0=0$. Будем считать, что полные заряды проводников известны и не могут изменяться. Заполним всё пространство между проводниками однородным простым диэлектриком с проницаемостью ε. Предположим, что распределение свободных зарядов на проводниках не изменилось. Докажем, что потенциал, создаваемый всеми зарядами в системе (свободными и связанными), удовлетворяет уравнению Лапласа. Так как выполняется (38), то создаваемый в диэлектрике также потенциал уменьшается в ε раз: $\varphi=\varphi_0/\varepsilon$. Очевидно, что при этом уравнение Лапласа выполняется: $\Delta\varphi=(\Delta\varphi_0)/\varepsilon=0$. Сохраняется и эквипотенциальность поверхности проводников. Это обосновывает сделанное предположение о неизменности распределения свободного заряда на проводниках после заполнения системы простым однородным диэлектриком.

В общем случае, учитывающем возможность плавного изменения ε в неоднородном диэлектрике, формула (60) позволяет утверждать, что "...электрическая индукция $\vec{D}(\vec{r})$ при заданных зарядах на проводниках...не зависит от диэлектрической проницаемости среды $\varepsilon(\vec{r})$...если последняя постоянна на эквипотенциальных поверхностях (т.е. $\nabla\varphi\times\nabla\varepsilon=0$)" ([28],

с.21). Действительно, условие $\nabla\varphi\times\nabla\varepsilon=0$ делает невозможным существование объёмного ФПТ. Поверхностного ФПТ тоже нет, так как $\vec{E} \parallel \vec{n}$ на границе диэлектрика с проводниками.

Изменяется ли распределение свободных зарядов по поверхности проводников при заполнении всего пространства неоднородным простым диэлектриком при условии $\nabla\varphi\times\nabla\varepsilon=0$ ? Предположим, что нет. Тогда не изменится, согласно (20), и поле вектора электрической индукции $\vec{D}_0(\vec{r})$, поскольку $\vec{j}_P=0$ и $\vec{i}_P=0$, и, кроме того, направление вектора $\vec{E}(\vec{r})=-\nabla\varphi$ не изменяется. Действительно, напряженность $\vec{E}(\vec{r})$ после добавления диэлектрика будет равна:

$$\vec{E}(\vec{r}) = \vec{D}_0(\vec{r})/(\varepsilon_0\varepsilon(\vec{r})) = \vec{E}_0(\vec{r})/\varepsilon(\vec{r}) = -\nabla\varphi_0/\varepsilon(\vec{r}), \qquad (62)$$

Как следствие, каждая эквипотенциальная поверхность (в том числе, поверхность проводников) сохранила это свойство. Кроме того, поле $\vec{E}(\vec{r})$ остаётся потенциальным:

$\text{rot}(\vec{E}(\vec{r}))=\text{rot}\{\vec{D}_0(\vec{r})/(\varepsilon_0\varepsilon(\vec{r}))\}=\nabla(1/\varepsilon(\vec{r}))\times(\vec{D}_0(\vec{r})/\varepsilon_0)+(1/\varepsilon(\vec{r}))\text{rot}\{\vec{D}_0(\vec{r})/\varepsilon_0\}=$

$=-\nabla\varepsilon\times\vec{D}_0(\vec{r})/[\varepsilon_0(\varepsilon(\vec{r}))^2]==-\nabla\varepsilon\times\nabla\varphi_0/(\varepsilon(\vec{r}))^2$

Так как $\nabla\varphi=-\vec{E}(\vec{r})=\nabla\varphi_0/\varepsilon(\vec{r})$, то $\text{rot}(\vec{E}(\vec{r}))=-\nabla\varepsilon\times\nabla\varphi/\varepsilon(\vec{r})=0$, т.е. поле $\vec{E}(\vec{r})$ действительно потенциальное. Так как при нашем предположении неизменности $\vec{D}_0(\vec{r})$ сохраняются полные заряды проводников и их эквипотенциальность, выполняются уравнения Максвелла $\text{div}(\vec{D})=0$ и $\text{rot}(\vec{E})=0$, найденное решение, предполагающее неизменность расположения свободных зарядов на проводниках, можно считать единственным решением электростатической задачи [29].

Рассмотрим ещё две стандартные задачи, в которых пространство между проводниками заполняется простым диэлектриком. В первой задаче $\nabla\varphi\times\nabla\varepsilon=0$ и поэтому $\vec{j}_P=0$ и $\vec{i}_P=0$. Во второй $\nabla\varphi\times\nabla\varepsilon\neq0$ и поэтому появляется ФПТ. В обеих задачах две тонкие концентрические проводящие сферы образуют конденсатор, заряд внутренней сферы $+Q$, внешней $-Q$. В первой задаче (рис.7) пространство между сферами заполнено простым неоднородным диэлектриком с проницаемостью $\varepsilon(r)$, зависящей только от расстояния от центра сфер.

Во второй задаче (рис.8) два различных простых однородных диэлектрика с проницаемостями $\varepsilon_1$ и $\varepsilon_2$ занимают по половине пространства между сферами. Пока между сферами нет диэлектрика, заряд распределён по их поверхности равномерно. В этом случае напряжённость поля внутри первой сферы и снаружи второй равна нулю, между сферами $\vec{E}=Q\vec{r}/(4\pi\varepsilon_0 r^3)$, $\varphi=Q/(4\pi\varepsilon_0 r)$, $\vec{D}=\varepsilon_0\vec{E}$. Во всех трёх областях справедливы уравнения $\text{rot}(\vec{E})=0$ и $\text{div}(\vec{D})=0$. Последнее уравнение эквивалентно (в рассматриваемой задаче) уравнению Лапласа для потенциала Δφ=0. В качестве граничных условий можно взять эквипотенциальность проводников и сохранение на них заряда (первое условие - дифференциальное, второе - интегральное). Это решение, угаданное благодаря симметрии, единственное [22].

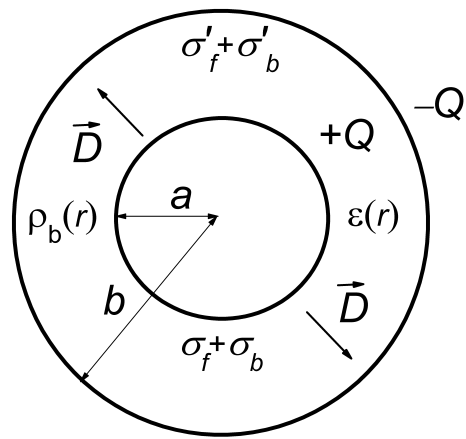

Рис.7 Неоднородный простой диэлектрик в сферическом конденсаторе. Q - полный свободный заряд обкладок, $\sigma_f$ и $\sigma_b$ ($\sigma_f'$ и $\sigma_b'$) - поверхностная плотность свободного и связанного заряда, соответственно, на границе малой (большой) проводящей сферы с диэлектриком. Диэлектрическая проницаемость $\varepsilon(r)$ зависит только от расстояния $r$ от центра. Линии $\vec{D}$ (и остальных полей) направлены радиально. Из-за неоднородности диэлектрика внутри него возникает объёмный связанный заряд с плотностью $\rho_b(r)$.

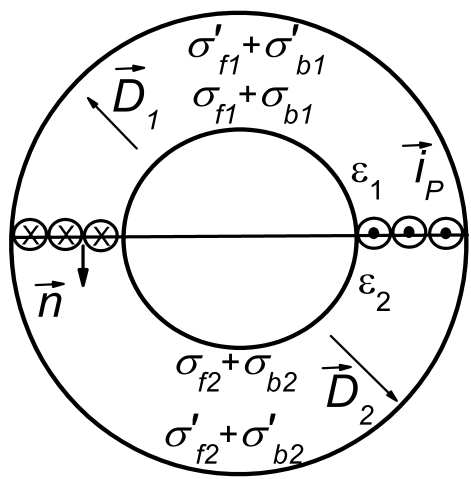

Рис.8 Система, подобная изображённой на рис.7, но диэлектрическая проницаемость различна в верхней и нижней половине сферического конденсатора ($\varepsilon_1$ и $\varepsilon_2$, соответственно), $\vec{n}$ – вектор нормали к границе диэлектриков. Обозначения поверхностной плотности зарядов такие же, как на рис.7, цифры "1" и "2" учитывают различия верхнего и нижнего диэлектриков. Линии $\vec{D}$ (и остальных полей) направлены радиально, но различны в верхней и нижней половине системы. Кружками показан ФПТ на границе диэлектриков, $\vec{i}_P$. – его линейная плотность.

При заполнении конденсатора изотропным диэлектриком, неоднородность которого не нарушает сферической симметрии, свободные заряды на проводниках сохраняют равномерное распределение. В пространстве между сферами, вследствие теоремы Остроградского-Гаусса, $\vec{D}=Q\vec{r}/(4\pi r^3)$ и, следовательно,

$\vec{E}=\vec{D}/(\varepsilon_0\varepsilon(r))$, $\varphi=Q/(4\pi\varepsilon_0\varepsilon(r)r)$, $\vec{P}=\varepsilon_0(\varepsilon(r)-1)\vec{E}$.

На границе внутренней сферы и диэлектрика возникает согласно (37) отрицательный связанный заряд $\sigma_b=-\sigma_f(\varepsilon(r)-1)/\varepsilon$, где $\sigma_f=Q/(\pi a^2)$ – плотность свободного заряда на внутренней сфере. На границе диэлектрика и внешней сферы возникает положительный связанный заряд $\sigma'_b=-\sigma'_f(\varepsilon(r)-1)/\varepsilon$, где $\sigma'_f=-Q/(\pi b^2)$ – плотность свободного заряда на внешней сфере. Суммарный связанный заряд на границах равен:

$Q_{b,гр}=-4\pi a^2\sigma_f(\varepsilon(a)-1)/\varepsilon(a)+4\pi b^2\sigma'_f(\varepsilon(b)-1)/\varepsilon(b)=4\pi(a^2\sigma_f/\varepsilon(a)-b^2\sigma'_f/\varepsilon(b))=Q(1/\varepsilon(a)-1/\varepsilon(b))$.

В пространстве между сферами возникает объёмный связанный заряд с плотностью:

$\rho_b=-\text{div}(\vec{P})=-\varepsilon_0\nabla\varepsilon(r)\cdot\vec{E}-\varepsilon_0(\varepsilon(r)-1)\text{div}(\vec{E})=-\varepsilon_0\nabla\varepsilon(r)\cdot\vec{E}-(\varepsilon(r)-1)\rho_b \Rightarrow$

$\rho_b=-\varepsilon_0\nabla\varepsilon(r)\cdot\vec{E}/\varepsilon(r)=(d\varepsilon(r)/dr)Q/(4\pi\varepsilon^2(r)r^2)$

где учтено $\text{div}(\vec{E})=\rho_b/\varepsilon_0$ и $\nabla\varepsilon=-(d\varepsilon(r)/dr)\vec{r}/r$. Суммарный объёмный связанный заряд равен: $Q_{b,об}=Q(1/\varepsilon(b)-1/\varepsilon(a))$. Таким образом, суммарный связанный заряд системы равен нулю: $Q_{b,гр}+Q_{b,об}=0$. ФПТ в рассматриваемой системе не возникает, так как $\text{rot}(\vec{P})=\varepsilon_0\nabla\varepsilon(r)\times\vec{E}+\varepsilon_0(\varepsilon(r)-1)\text{rot}(\vec{E})=0$.

Итак, если в простом неоднородном диэлектрике, заполняющем пространство между проводниками, вектор $\nabla\varepsilon(r)$ параллелен линиям поля, ФПТ отсутствует.

Для второй задачи, в которой два простых однородных диэлектрика образуют границу в экваториальной плоскости конденсатора (рис.8), единственное решение соответствует неравномерному распределению свободных зарядов на внутренней и внешней сферах, при котором, однако, полный заряд распределен равномерно. Это решение удовлетворяет (и поэтому является единственным), как уравнениям $\text{rot}(\vec{E})=0$ и $\text{div}(\vec{D})=0$, так и граничным условиям. Пусть $\varepsilon_2>\varepsilon_1$ (см. рис.8). Тогда плотности свободного заряда $\sigma_{f1}$ и $\sigma'_{f1}$ и для полусфер, контактирующих с диэлектриком с меньшей проницаемостью, будут меньше соответствующих $\sigma_{f2}$ и $\sigma'_{f2}$ плотностей свободного заряда, для полусфер, контактирующих с диэлектриком с большей проницаемостью. При этом, плотность связанных зарядов компенсирует эту разность, так, что $\sigma_{f1}+\sigma_{b1}=\sigma_{f2}+\sigma_{b2}$ и $\sigma'_{f1}+\sigma'_{b1}=\sigma'_{f2}+\sigma'_{b2}$. Кроме того, согласно (37), $\sigma_{b1}=-\sigma_{f1}(\varepsilon_1-1)/\varepsilon_1$, $\sigma_{b2}=-\sigma_{f2}(\varepsilon_2-1)/\varepsilon_2$, $\sigma'_{b1}=-\sigma'_{f1}(\varepsilon_1-1)/\varepsilon_1$, $\sigma'_{b2}=-\sigma'_{f2}(\varepsilon_2-1)/\varepsilon_2$ и, согласно, закону сохранения заряда $Q=2\pi a^2(\sigma_{f1}+\sigma_{f2})$ и $-Q=2\pi b^2(\sigma'_{f1}+\sigma'_{f2})$. Из этих уравнений следует:

$\sigma_{f1}=Q\varepsilon_1/\{2\pi a^2(\varepsilon_1+\varepsilon_2)\}$, $\sigma'_{f1}=-Q\varepsilon_1/\{2\pi b^2(\varepsilon_1+\varepsilon_2)\}$ и $\sigma_{f2}=Q\varepsilon_2/\{2\pi a^2(\varepsilon_1+\varepsilon_2)\}$, $\sigma'_{f2}=-Q\varepsilon_2/\{2\pi b^2(\varepsilon_1+\varepsilon_2)\}$. Из теоремы Остроградского-Гаусса можно найти напряжённость поля в обоих диэлектриках в пространстве между сферами: $\vec{E}=Q\vec{r}/(2\pi\varepsilon_0(\varepsilon_1+\varepsilon_2)r^3)$. Зная напряжённость, можно определить однонаправленные с вектором $\vec{E}$ электрическую индукцию и поляризацию: $\vec{D}_1=\varepsilon_0\varepsilon_1\vec{E}$, $\vec{D}_2=\varepsilon_0\varepsilon_2\vec{E}$ и $\vec{P}_1=\varepsilon_0(\varepsilon_1-1)\vec{E}$, $\vec{P}_2=\varepsilon_0(\varepsilon_2-1)\vec{E}$.

По формуле (61) можно найти линейную плотность ФПТ (кружки на рис.8):
$$\vec{i}_P=(\varepsilon_2-\varepsilon_1)Q\vec{n}\times\vec{r}/(2\pi\varepsilon_0(\varepsilon_1+\varepsilon_2)r^3),$$
где $\vec{n}$ - как и ранее единичный вектор нормали, ориентированный из первой среды во вторую. Видно, что модуль линейной плотности кольцевого ФПТ уменьшается по мере удаления от центра сфер как $\sim 1/r^2$.

Второй пример иллюстрирует правило - в неоднородном простом диэлектрике, заполняющем пространство между проводниками, возникает ФПТ, если вектор $\nabla\varepsilon(r)$ не параллелен линиям поля. В рассмотренной задаче градиент проницаемости возникает на границе диэлектриков и направлен перпендикулярно линиям поля.

Итак, проще всего задача нахождения статического поля $\vec{D}$ по его источникам решается тогда, когда присутствует только один вид источников. Это, например, свободные заряды на проводниках в однородном простом диэлектрике и связанные заряды в системе диэлектрических тел с "замороженной" поляризацией в вакууме. В первом случае задача решается как для тех же свободных зарядов в вакууме с заменой $\varepsilon_0\to\varepsilon_0\varepsilon$. Во втором случае задача может решаться двумя способами: с помощью формулы (15) с зарядами (7) и (39) или по формуле (20) с ФПТ (17) и (21). Первый способ предполагает непосредственное нахождение вектора $\vec{E}$, после чего, если это необходимо, с помощью (33) может быть найден $\vec{D}$. Второй способ предпочтителен, если известно решение аналогичной магнитостатической задачи [8].

## 7. Токи проводимости и намагничивания в магнитостатике

Как отмечалось в разделе 5, в магнитостатических задачах обычно считаются заданными токи проводимости и ФМЗ, если в задаче присутствуют постоянные магниты. Для нахождения токов намагничивания часто используют условия их связи с токами проводимости. Так, для однородного простого магнетика из уравнений (9), (10), (30), (31) следует связь токов намагничивания и проводимости:

$$\vec{j}_M = (\mu-1)\vec{j}_c \qquad (63)$$

а также полного тока и тока проводимости:

$$\vec{j} = \vec{j}_M + \vec{j}_c = \mu \vec{j}_c \qquad (64)$$

Отметим, что соотношения (63), (64) можно применять только для линейных токов проводимости, в частности, для токов в проводах бесконечно малого сечения. Математически это связано с локальностью уравнений (30) и (31). С физической точки зрения, это обусловлено тем, что в формулу (31) входит напряжённость магнитного поля, зависящая в общем случае, согласно (26), от ФМЗ. Вкладом в напряжённость от ФМЗ можно пренебречь, если расстояние $r$ от проводника до области в магнетике, где протекает ток намагничивания, стремится к нулю и поэтому собственный вклад тока в проводнике в напряженность неограниченно возрастает по закону $\sim 1/r$, и вкладами в $\vec{H}$ всех остальных источников можно пренебречь. В этом случае ток намагничивания в (63) на границе линейного проводника с внешней средой зависит только от тока проводимости.

Из (25) и (64) следует, что система замкнутых неподвижных линейных токов в однородном простом магнетике с проницаемостью $\mu$, заполняющем всё пространство, создаёт индукцию магнитного поля в $\mu$ раз большую, чем создают те же токи проводимости в вакууме. Поэтому утверждения (У3а) и (У3б) справедливы, но с уточнением − токи должны быть линейными. Заметим, что утверждение (У3а), более сложное, чем (У3б), так как делает заключение относительно силы, действующей на проводник с током, но для линейных токов оно эквивалентно (У3б).

Токи намагничивания могут возникнуть на любой границе. Если граница разделяет простые магнетики с проницаемостью $\mu_1$ и $\mu_2$, соответственно, то граничное условие для тангенциальной компоненты вектора $\vec{H}$ даст соотношение:

$B_{2\tau}/(\mu_0\mu_2) − B_{1\tau}/(\mu_0\mu_1) = i_с \Rightarrow$

$(B_\tau^{вн}/\mu_0 + (i_М + i_с)/2)/\mu_2 − (B_\tau^{вн}/\mu_0 − (i_М + i_с)/2)/\mu_1 = i_с$ \qquad (65)

где $B_{\tau 1}$ и $B_{\tau 2}$ − проекция на касательную к границе индукции магнитного поля в первой и второй среде, соответственно (в точках, бесконечно близких к границе), $B_\tau^{вн} = (B_{\tau 1} + B_{\tau 2})/2$ − тангенциальная компонента индукции внешнего поля в рассматриваемой точке границы, $i_М$ и $i_с$ − линейные плотности поверхностных токов намагничивания и проводимости, соответственно. Под "внешним" понимается поле, создаваемое всеми токами, кроме токов проводимости $i_с$ и токов намагничивания $i_М$ на малой площадке в рассматриваемой точке границы.

Из (65) следуют формулы для линейной плотности поверхностного тока намагничивания:

$i_М = i_с(2\mu_1\mu_2/(\mu_1 + \mu_2) − 1) + 2B_\tau^{вн}(\mu_2 − \mu_1)/\mu_0(\mu_1 + \mu_2)$ \qquad (66)

и для полного тока:

$i = i_C + i_M = 2i_c \mu_1 \mu_2 / (\mu_1 + \mu_2) + 2B_\tau^{\text{вн}}(\mu_2 - \mu_1)/\mu_0(\mu_1 + \mu_2)$ (67)

Из (66) видно, что на границе раздела простых магнетиков могут существовать токи намагничивания, даже если на этой границе отсутствуют токи проводимости ($i_c = 0$).

Если среда 1 представляет собой постоянный магнит, граничное условие для тангенциальной проекции вектора $\vec{H}$ запишется как:

$B_{2\tau}/(\mu_0 \mu_2) - H_{1\tau} = i_c \Rightarrow$

$(B_{2\tau}^{\text{вн}}/\mu_0 + (i_M + i_c)/2)/\mu_2 - (H_{1\tau}^{\text{вн}} - i_c/2)/\mu_1 = i_c$ (68)

где $B_{2\tau}$ и $H_{1\tau}$ – проекция на касательную к границе индукции и напряженности магнитного поля, соответственно, во второй и первой среде (в точках, бесконечно близких к границе), $B_{2\tau}^{\text{вн}}$ и $H_{1\tau}^{\text{вн}}$ соответствующие вклады внешних полей в тангенциальные компоненты векторов магнитной индукции и напряженности магнитного поля, соответственно, в рассматриваемой точке поверхности. В формулах (65)-(68) знаки токов определяются по правилу "правого винта". В (68) учтено, что ФМЗ рассматриваемого малого участка границы не даёт вклад в циркуляцию при выводе граничного условия для тангенциальной проекции вектора $\vec{H}$. Отметим, что величину $H_{1\tau}^{\text{вн}}$ можно, при необходимости, выразить через тангенциальные проекции индукции внешнего поля $B_{1\tau}^{\text{вн}}$ и намагниченности $M_{1\tau}$ в первой среде. Вектор намагниченности обычно считается известным в таких задачах. При этом можно использовать связь между проекциями векторов в (34) на касательную к границе: $H_{1\tau} = B_{1\tau}/\mu_0 - M_{1\tau}$, откуда следует $H_{1\tau}^{\text{вн}} = B_{1\tau}^{\text{вн}}/\mu_0 - i_M/2 - M_{1\tau}$.

Расчёт токов намагничивания по формулам (66), (68) в общем случае затруднён, в частности, тем, что они содержат компоненты полей ($B_{1\tau}^{\text{вн}}$, $B_{2\tau}^{\text{вн}}$, $B_\tau^{\text{вн}}$), которые могут быть рассчитаны, только если известны все токи. Т.е. возникает та же проблема, которая отмечалась при обсуждении электростатического уравнения (47).

Для симметричных систем уравнения (66), (67) могут упрощаться. Например, пусть ток проводимости с линейной плотностью $i_c$ течет вдоль тонкостенной цилиндрической длинной трубы. Внутри трубы - вакуум, всё пространство снаружи заполнено простым однородным магнетиком с магнитной проницаемостью $\mu$. Таким образом, система характеризуется цилиндрической симметрией. Тогда (66) запишется так:

$i_M = i_c(\mu - 1)/(1 + \mu) + 2B_\tau^{\text{вн}}(\mu - 1)/[\mu_0(1 + \mu)]$

Рассчитывая индукцию магнитного поля внутри и снаружи трубы с помощью теоремы о циркуляции, получаем:

$B_\tau^{\text{вн}} = (B_{\tau 1} + B_{\tau 2})/2 = (i_M + i_c)/2$.

Поэтому $i_M = i_c(\mu - 1)$, что аналогично (66) с заменой объёмной плотности тока на линейную.

В рассмотренном примере равны нулю нормальные компоненты магнитных векторов на границе сред, поэтому, согласно (28), ФМЗ отсутствует. В следующем разделе оба вида источников $\vec{H}$ будут рассмотрены более подробно.

## 8. Два вида источников поля $\vec{H}$: токи проводимости и "фиктивные магнитные заряды"

Вектор напряжённости магнитного поля $\vec{H}$ часто называется "вспомогательным", как и вектор электрической индукции $\vec{D}$. Из (34) следует тождество, служащее определением $\vec{H}$:

$$\vec{H} = \vec{B}/\mu_0 - \vec{M} \qquad (69)$$

Необходимость введения вектора $\vec{H}$, как правило, обосновывается уравнением (9), согласно которому ротор $\vec{H}$ зависит только от токов проводимости, и, значит, для симметричных систем вектор $\vec{H}$ может быть вычислен, даже если токи намагничивания неизвестны. Например, в длинном соленоиде, полностью заполненном простым магнетиком, вектор $\vec{H}$ зависит только от силы тока проводимости в обмотке, но не зависит от намагниченности вещества. Точнее, влиянием намагниченности можно пренебречь в центральной части соленоида, удалённой от его концов (т.е. в модели "длинного соленоида", когда пренебрегают краевыми эффектами). В этом случае для расчёта $\vec{H}$ по формуле (26) следует использовать только два первых слагаемых.

В первой половине двадцатого века существенно изменился подход в изложении магнитостатики: вместо магнитных зарядов (магнитных полюсов) постоянных магнитов основным объектом теории стали токи намагничивания. Поэтому второй источник вектора $\vec{H}$ – ФМЗ, либо вообще не упоминается в современной учебной литературе ([2], [4], [5], [20]), либо кратко обсуждается в теме, посвящённой постоянным магнитам ([18], [25]), и лишь в некоторых учебниках (например, [16], [23], [31], [32]) можно найти достаточно подробное описание всех видов источников $\vec{H}$.

Именно благодаря понятию ФМЗ вектор $\vec{H}$ особенно удобен при расчёте полей постоянных магнитов. Как следует из (26), в этом случае, а также и в других, когда отсутствуют токи проводимости, напряженность магнитного поля рассчитывается по закону Кулона, в котором вместо реальных зарядов нужно использовать ФМЗ.

Рассмотрим, например, тороидальный постоянный магнит с намагниченностью $\vec{M}$, линии которой представляют окружности с центрами на оси симметрии тороида (рис.4). Намагниченность $\vec{M}$ внутри постоянного тороидального магнита может быть создана полем тока проводимости в длинном проводе, направленном вдоль оси тороида, и, следовательно, описывается формулой (ср. (10.29) в [31], с.308):

$$\vec{M} = \frac{C}{r}\vec{\theta}_u \qquad (70)$$

где $C$ - постоянная, $\vec{\theta}_u$ - единичный вектор, касательный к линии поля (см.рис.4).

Если в магните сделать узкий зазор (рис.6), перпендикулярный линиям $\vec{M}$, на его плоских границах появится ФМЗ, кулоновское поле которого, согласно (26), определит величину $\vec{H}$ в зазоре. Задача расчёта $\vec{H}$ упрощается, если применима модель "плоского конденсатора", когда тороид тонкий, а зазор узкий, т.е. $(a+b)/2 >> (b-a) >> d$ (см. рис.6). Если нарушается первое из указанных сильных неравенств, плоские поверхности зазора нельзя считать параллельными (см. рис.8.17 в [31], с.255).

Из формулы (24) следует, что объёмные ФМЗ отсутствует, если div($\vec{M}$)=0. Это условия выполняется, в частности, когда намагниченность однородна. В случае тороида на рис.6, намагниченность неоднородна, но, тем не менее, div($\vec{M}$)=div($C\vec{\theta}_u/r$)=0, что согласуется с вихревым характером поля прямого провода, намагнитившего материал тороида. Поэтому внутри постоянного магнита на рис.6 объёмный ФМЗ отсутствует.

Однородность намагниченности, как правило, нарушается на границе сред с разными магнитными свойствами, что может приводить к появлению поверхностного ФМЗ. В общем случае можно написать граничное условие, эквивалентное (28):

$M_{2n} - M_{1n} = -\sigma_M \qquad (70)$

где $\sigma_M$ - поверхностная плотность фиктивных магнитных зарядов.

Согласно (70), на плоских границах узкого зазора в тороидальном магните (рис.6) поверхностная плотность ФМЗ $\sigma_M$ равна $M$. В модели идеального "плоского конденсатора" заряды на "обкладках" с поверхностной плотностью $\sigma_M$ создают напряженность в зазоре $H_g = M$. Этот результат может быть получен также из условия непрерывности нормальной компоненты $\vec{B}$ и уравнения (31) с учётом $\mu=1$. Таким образом, если постоянный магнит "абсолютно жёсткий", т.е. его намагниченность не зависит от внешнего поля, напряжённость магнитного поля в зазоре не зависит от радиуса тороида вопреки широко распространённому ошибочному мнению (см. ниже).

В модели "абсолютно жёсткого" магнита достаточно просто учитываются конечные размеры зазора. Для этого подобно электростатическому случаю (см. рис.5) можно записать индукцию магнитного поля системы на рис.6 в виде суммы поля $\vec{B}_0$, создаваемого тороидом без зазора (рис.4), и поля $\vec{B}'$ тонкой катушки, мысленно помещённой в зазор так, что токи (намагничивания) катушки и тороида компенсируют друг друга на боковой поверхности зазора. Линейная плотность токов намагничивания равна $M$.

Так же, как в электростатической задаче, рассмотренной выше, $|\vec{B}'| << |\vec{B}_0| = |\mu_0 \vec{M}|$. Напряжённость магнитного поля снаружи зазора равна $\vec{H}' = \vec{B}'/\mu_0$, причем эта формула справедлива и внутри тороида: $\vec{H} = \vec{B}/\mu_0 - \vec{M}/\varepsilon_0 = (\vec{B}_0 + \vec{B}')/\mu_0 - \vec{M} = \vec{B}'/\mu_0 = \vec{H}'$, так как $\vec{B}_0 = \mu_0 \vec{M}$.

Поле $\vec{B}'$ тонкой катушки на её оси можно найти с помощью общей формулы для поля на оси однослойного цилиндрического соленоида конечной длины (см. например, (10.12) в [32], с.168):

$$B = \frac{\mu_0 n I}{2}(\cos\alpha_1 - \cos\alpha_2) \qquad (71)$$

где $n$ - плотность витков обмотки соленоида, $I$ - сила тока в обмотке, $\alpha_1$ и $\alpha_2$ - углы, под которыми из точки, где ищется поле, видны радиусы торцов соленоида.

Величина индукции в точке на оси тороида на рис.6, вблизи границы с зазором (внутри магнетика), найдётся из (71) с учётом $nI=M$, $\cos(\alpha_1)=d/(d^2+R^2)^{1/2}$, $\cos(\alpha_2)=0$, где $R=(b-a)/2$ - радиус торца зазора и, следовательно, радиус "катушки":

$$B = B_0 + B' = \mu_0 M (1 - \frac{d}{2\sqrt{R^2+d^2}}) \approx \mu_0 M(1 - \frac{d}{2R}) \qquad (72)$$

где учтена малость зазора ($d<<R$). На расстояниях от зазора много больших $R$ магнитная индукция поля "катушки" $\vec{B}'$ описывается формулой поля магнитного диполя, т.е. убывает обратно пропорционально кубу расстояния.

Из непрерывности нормальной компоненты вектора $\vec{B}$ следует, что напряжённость магнитного поля на оси тороида внутри тонкого зазора равна:

$$H_g \approx M(1 - \frac{d}{2R}) \qquad (73)$$

Снаружи зазора напряжённость магнитного поля быстро убывает с расстоянием, её абсолютное значение в точке, для которой записана формула (72), можно получить из (69) и (72):

$$|\vec{H}'| \approx \frac{Md}{2R} \approx H_g \frac{d}{2R} \qquad (74)$$

Вектор $\vec{H}'$ в (74) направлен противоположно намагниченности, поэтому его можно назвать размагничивающим полем. Из (74) следует, что величина размагничивающего поля пропорциональна ширине зазора.

Экспериментальные данные [32] свидетельствуют, что для постоянных магнитов из стали и железа, намагниченность которых заметно зависит от напряжённости локального поля, размагничивающее поле заметно уменьшает намагниченность материала магнита вблизи зазора. При этом, в согласии с формулой (73), уменьшается и индукция магнитного поля внутри зазора. Измерения с помощью баллистической катушки [32] показали, что максимальное магнитной значение индукции (и намагниченности материала) достигается в сечении тороида, лежащем диаметрально противоположно разрезу. Это согласуется с минимальным значением размагничивающего поля в этом сечении.

Экспериментально наблюдаемое уменьшении магнитной индукции в зазоре (по сравнению с остальными участками тороида) можно формально учесть [32] с помощью эффективного сечения магнитного потока $S=n\pi R^2$, где $n=1$ для удалённых от зазора участков и $n>1$ для поля внутри зазора, которое можно найти из условия $BS$=const. Эксперимент даёт зависимость параметра $n$ от ширины и радиуса зазора в виде:

$$n = 1 + a\frac{d}{R} \qquad (75)$$

где $a$ - численный коэффициент, равный $\approx 7$ для стальных и железных постоянных магнитов [32]. Формула (75), как и (73), (74), указывает на отношение ($d/R$) как основной параметр, влияющий на размагничивающее поле и магнитное поле в зазоре.

Задача с узкой щелью в тороидальном магните в том виде, как её можно встретить в подавляющем большинстве учебников (как отечественных, так и зарубежных), может служить иллюстрацией к актуальности совета комиссии "по закону Кулона" (см. раздел 2) чётко указывать область применимости формул и, в общем случае, используемой модели. К примеру, в задаче 2.303 из [30], требуется найти "модуль напряжённость магнитного поля внутри магнита" в системе, изображённой на рис.6. Почти во всех учебниках (ограничимся ссылкой на [5], с.184) решение этой задачи, основывается на теореме о циркуляции напряжённости магнитного поля, которую можно получить из (9) с использованием теоремы Стокса:

$$\oint \vec{H}d\vec{l} = \int_G \vec{H}d\vec{l} + \int_M \vec{H}d\vec{l} = 0 \qquad (76)$$

Формула (76) учитывает отсутствие токов проводимости. Интегрирование в (76) ведётся по замкнутому круговому контуру, лежащему на оси тороида, состоящему из двух частей: "G" - внутри зазора и "M" - внутри магнетика. Если зазор тонкий, то оправдана замена $\int_G \vec{H}d\vec{l}$ на произведение $H_g d$, где $H_g$ - почти однородное поле внутри зазора. Однако вне зазора напряжённость магнитного поля нельзя считать однородной даже приблизительно. Тем не менее в [5] и в подавляющем большинстве других книг по электромагнетизму интеграл $\int_M \vec{H}d\vec{l}$ заменяется на $H_M\pi(a+b)$, и таким образом вычисляется величина, называемая "напряжённостью магнитного поля внутри магнетика":

$H_M = -H_g d/\pi(a+b)$ (77)

в то время как $H_M$ из (77) представляет собой всего лишь среднее значение проекции напряжённости на окружность с радиусом $a+b$. В результате у читателя создаётся ложное представление о направлении и величине поля размагничивания $\vec{H}'$.

Иногда (76) используют даже для "расчёта" напряжённости поля внутри зазора [34]:

$H_g = -H_M\pi(a+b)/d$ (78)

иллюстрируя увеличение напряжённости магнитного поля в зазоре при уменьшении его ширины $d$. Так как $H_M$ само зависит от $d$, использование (78) для этой цели малоубедительно.

Если намагниченность тороидального магнита практически не зависит от внешнего поля (этим свойством обладают современные редкоземельные магниты), вектор $\vec{H}'$ можно вычислить в произвольной точке, используя, например, решение эквивалентной задачи о напряжённости электрического поля снаружи плоского конденсатора [35]. Если намагниченность зависит от внешнего поля, то точный расчёт $\vec{H}'$ усложняется, так как намагниченность становится неоднородной и, в частности, линии $\vec{M}$ нельзя уже считать окружностями.

Как отмечалось в разделе 2, напряжённость магнитного поля постоянного магнита, помещённого в простой магнетик, уменьшается по сравнению полем этого магнита в вакууме в общем случае не в $\mu$ раз, где $\mu$ - магнитная проницаемость магнетика. Дж.К.Максвелл рассмотрел [13] этот вопрос только для случая бесконечно тонких длинных продольно намагниченных постоянных магнитов, концы которых могут рассматриваться как "магнитные полюса" - аналог электрических точечных зарядов. Напряжённость магнитного поля магнитных полюсов уменьшается в простом магнетике в $\mu$ раз так же, как напряжённость точечных электрических зарядов уменьшается в простом диэлектрике в $\varepsilon$ раз. Однако, если постоянный магнит, погружённый в безграничный простой магнетик, имеет конечные размеры, создаваемая им напряжённость магнитного поля зависит от его формы. В работе [10] решение уравнения Лапласа для постоянных сфероидальных магнитов в простом магнетике показало, в частности, уменьшение напряжённости (по сравнением с вакуумом) для сферического магнита в $(2\mu+1)/3$ раз, а для магнитов в виде "иголок" в $\mu$ раз.

При решении уравнения Лапласа в [10] использовались граничные условия, в частности, условие для нормальной компоненты напряжённости магнитного поля на границе двух простых магнетиков, которое мы запишем в виде, аналогичном (43):

$\mu_2\mu_0 H_{n2} - \mu_1\mu_0 H_{n1} = 0 \Rightarrow$

$\mu_2(H_n^{\text{вн}} + \sigma_M/2) - \mu_1(H_n^{\text{вн}} - \sigma_M/2) = 0$          (79)

где $H_n^{\text{вн}} = (H_{n1} + H_{n2})/2$ – напряженность магнитного поля в точке на границе в середине рассматриваемого бесконечно+ малого участка поверхности.

Из (79) следует формула для поверхностной плотности "фиктивных магнитных зарядов", выраженной через нормальную проекцию напряжённости внешнего магнитного поля в точке границы двух простых магнетиков:

$\sigma_M = 2 H_n^{\text{вн}} (\mu_1 - \mu_2)/(\mu_1 + \mu_2)$          (80)

Например, в задаче 2.298 из [30] уравнение (80) описывает плотность ФМЗ на плоской границе раздела двух полупространств, заполненных двумя различными однородными простыми магнетиками. В (80) следует подставить напряжённость $H_n^{\text{вн}} = I/(2\pi r)$ поля, создаваемого линейным прямым током проводимости $I$ на границе магнетиков ($r$ - расстояние от тока проводимости до рассматриваемой точки границы). Суммарная напряжённость магнитного поля в этой задаче, равная сумме вкладов от тока проводимости и ФМЗ на границе раздела, выглядит своеобразно - линии поля $\vec{H}$ в каждом магнетике представляют собой окружности, терпящие разрыв на границе магнетиков. Величина $\vec{H}$ обратно пропорциональна магнитной проницаемости, поскольку поле вектора $\vec{B}$ симметрично, что можно доказать, решая задачу методом изображений для линейного провода, расположенного в одной из двух сред, и устремляя расстояние между проводом и границей к нулю.

Для нормальной компоненты напряжённости магнитного поля на границе постоянного магнита (среда 1) и простого магнетика (среда 2) можно записать граничное условие, аналогичное (52):

$\mu_2\mu_0 H_{2n} - B_{1n} = 0 \Rightarrow$

$\mu_2\mu_0 (H_{2n}^{\text{вн}} + \sigma_M/2) - B_{1n} = 0$          (81)

где $H_{2n}^{\text{вн}}$ учитывает все вклады в напряженность магнитного поля, кроме вклада ФМЗ, расположенных на рассматриваемом бесконечное малом участке поверхности. Токи проводимости, даже если они присутствуют на этом участке, вклада в нормальную компоненту напряжённости не дают.

Одним из отличий формул (79) и (81) от электростатических аналогов (43) и (52) является отсутствие в них плотности "магнитных свободных зарядов", в природе не обнаруженных.

В качестве примера использования (81) найдём напряжённость магнитного поля вблизи торца длинного ("полубесконечного") продольно намагниченного постоянного магнита ($\vec{M}$ =const) с геометрией рис.2 и заменой на нём $\sigma_b' \to \sigma_M'$, $\sigma_b'' \to \sigma_M''$, $\sigma_b \to 0$, $\varepsilon \to \mu$ и $\vec{P} \to \vec{M}$. Используя (34), запишем $B_{1n}$ в виде:

$B_{1n} = \mu_0(M + H_{1n}) = \mu_0(M - \sigma_M/2)$ (82)

Учитывая $H_{2n}^{вн} \ll \sigma_M/2$ и заменяя в (81) $\mu_2 \to \mu$, из (81) и (82) получаем:

$\sigma_M = 2M/(\mu+1)$ (83)

и

$H_{2n} = \sigma_M/2 = M/(\mu+1)$ (84)

аналогичные электростатическим формулам (57) и (58), соответственно.

Так же, как в электростатическом случае (см конец раздела 5), из (81) следует, что однородно намагниченная (перпендикулярно своей плоскости) тонкая пластина толщины $d$ не создаёт магнитного поля в окружающем её магнетике при $d \to 0$.

В неоднородной среде нужно учитывать возможность появления объёмного ФМЗ. Для неоднородного простого магнетика можно записать:

$\rho_M = -\text{div}(\vec{M}) = -\text{div}((\mu-1)\vec{B}/\mu_0\mu) = \nabla(1/\mu) \cdot (\vec{B}/\mu_0) - (\mu-1)\text{div}(\vec{B})/(\mu_0\mu) =$

$= -\nabla\mu \cdot (\vec{B}/\mu_0\mu^2) = -\nabla\mu \cdot \vec{H}/\mu$ (85)

Из (85) следует, что в неоднородном простом магнетике объёмный ФМЗ определяется скалярным произведением градиента магнитной восприимчивости и вектора напряжённости магнитного поля. Если проницаемость не изменяется вдоль линий магнитного поля, ФМЗ отсутствует. В последнем примере раздела 6 простой магнетик, расположенный вне длинного цилиндрического проводника с током, предполагался однородным. Если магнетик неоднородный, но проницаемость $\mu(r)$ зависит только от расстояния от оси цилиндра, то согласно (85) плотность ФМЗ равна нулю и модуль напряжённости магнитного поля вне цилиндра может быть выражен с помощью закона полного тока (теоремы о циркуляции) через ток проводимости $I$: $H = I/((2\pi r)$. Модуль индукции магнитного поля, следовательно, равен $B = \mu_0\mu(r)H = \mu_0\mu(r)I/((2\pi r)$.

Одним из видов неоднородности является граница между частями системы с различными магнитными свойствами. Такой границей может быть поверхность проводника, размеры которого нельзя считать пренебрежимо малыми. Например, если два цилиндрических проводника конечного радиуса (см. рисунок в [4] на с.254), помещены в однородный простой магнетик, то на границе каждого проводника вектор напряжённости поля, создаваемого соседним проводником, имеет ненулевую нормальную компоненту, которая согласно (85) вызовет появление на этой границе ФМЗ. В результате, в формуле (26) для напряжённости магнитного поля нужно будет учитывать все слагаемые, а утверждения (У3а) и (У3б) будут несправедливыми.

В упомянутой выше задаче 2.298 из [30] замена линейного тока проводимости на провод конечного радиуса с равномерным распределением тока проводимости по сечению значительно усложняет решение. Во-первых, уже нельзя пользоваться методом изображений и, во-вторых, необходимо учесть с помощью формулы (80) ФМЗ на цилиндрических границах "толстого" провода с магнетиками, возникающий благодаря нормальной компоненте $H_\text{n}^\text{вн}$ поля ФМЗ, расположенного на плоской границе раздела магнетиков. Для того, чтобы поле $\vec{B}$ в модифицированной задаче осталось симметричным, нужно сделать провод составным так, чтобы плотность тока проводимости в каждой из полуцилиндрических половинок была различна, а полный ток ($\vec{j} = \vec{j}_c + \vec{j}_M$) обладал цилиндрической симметрией. Тогда, аналогично электростатической задаче на рис.8, в которой перераспределение свободных зарядов по поверхности сферы обеспечило симметричное распределение полного заряда и сферическую симметрию поля $\vec{E}$, в рассматриваемой задаче симметричное распределение полного тока обеспечит, согласно (25) цилиндрическую симметрию поля $\vec{B}$. В отличие от свободных зарядов в электростатике, распределение которых на поверхности проводников может изменяться в зависимости от внешних условий, токи проводимости в магнитостатике обычно считаются заданными и их нужное распределение по сечению провода следует объяснить в условии задачи, например, подключением половинок провода к различным источникам электродвижущей силы.

Итак, проще всего задача нахождения статического поля $\vec{H}$ по его источникам решается тогда, когда присутствует только один вид источников. Это, например, линейные токи в однородном простом магнетике и ФМЗ в системе постоянных магнитов в вакууме. В первом случае задача решается как для тех же линейных токов в вакууме с заменой $\mu_0 \rightarrow \mu_0\mu$. Во втором случае задача сводится к электростатической с ФМЗ, определяемым по формулам (24) и (28). В остальных случаях основной трудностью, как правило, является учёт индуцированного ФМЗ.

## 9. Заключительные замечания

Теорему Гельмгольца и получаемые на её основе формулы для расчёта электростатических и магнитостатических полей целесообразно использовать в разделе

"Электромагнетизм" курса общей физики для студентов физических и технических специальностей. В результате у студентов возникнет цельное, без существенных пробелов, представление об источниках этих полей и их взаимосвязи. Можно предложить следующее краткое изложение сути формул (15), (20), (25), (26). (1) Источниками электростатических полей $\vec{E}$, $\vec{D}$ и магнитостатических полей $\vec{B}$, $\vec{H}$ являются заряды и токи. (2) Для "настоящих" полей $\vec{E}$ и $\vec{B}$ источники всегда реальны (заряды для $\vec{E}$ и токи для $\vec{B}$) и однотипны (либо заряд, либо ток). (3) "Вспомогательные" поля $\vec{D}$ и $\vec{H}$ могут создаваться как реальными источниками (свободными зарядами для $\vec{D}$ и токами проводимости для $\vec{H}$), так и "фиктивными" ("поляризационными токами" для $\vec{D}$ и "магнитными зарядами" для $\vec{H}$). Источники $\vec{D}$ и $\vec{H}$ разнотипны как с точки зрения свойства "реальность-фиктивность", так и свойства "заряд-ток". (4) Связь между источником и полем всегда описывается либо законом Кулона, если источник – заряд, либо законом Био-Савара-Лапласа, если источник – ток.

Формулы (15), (20), (25), (26) убедительно обосновывают необходимость различать "настоящие" и "вспомогательные" поля, так как первые создаются только реальными источниками, а вторые – реальными и "фиктивными". Кроме того, симметричное строение этих формул, вместе с парными формулами (7) и (24), (10) и (17), подсказывает, как использовать аналогию между математическими формулами в электростатических и магнитостатических задачах. Ещё одно немаловажное достоинство теоремы Гельмгольца в том, что она позволяет полнее оценить роль уравнений Максвелла. Действительно, формула (1) становится практически полезной, только если известны ротор и дивергенция векторной величины. Отметим, что разобранные в статье иллюстративные задачи не требуют для своего решения применения уравнений Лапласа или Пуассона (достаточно использовать граничные условия).